\documentclass[runningheads]{llncs}

\usepackage{eccv}

\usepackage{eccvabbrv}

\usepackage{graphicx}
\usepackage{booktabs}

\usepackage[accsupp]{axessibility}  

\usepackage{hyperref}

\usepackage{orcidlink}

\usepackage{colortbl}
\usepackage{xcolor}
\usepackage{tcolorbox}
\usepackage[dvipsnames]{xcolor}

\usepackage{xcolor}
\usepackage{soul}
\usepackage{wrapfig}
\usepackage{fontawesome5}

\definecolor{mygreen}{rgb}{0.286,0.737,0.392}
\definecolor{bblue}{HTML}{4F81BD}
\definecolor{oorange}{HTML}{F4C842}
\definecolor{rred}{HTML}{C0504D}
\definecolor{ggreen}{HTML}{9BBB59}
\definecolor{ppurple}{HTML}{9F4C7C}
\definecolor{darkgreen}{HTML}{228B22}
\definecolor{cred}{HTML}{D81B60}
\definecolor{cblue}{HTML}{1E88E5}
\definecolor{cyellow}{HTML}{FFC107}
\definecolor{nred}{HTML}{e41a1c}
\definecolor{nblue}{HTML}{377eb8}
\definecolor{ngreen}{HTML}{4daf4a}
\definecolor{lblue}{HTML}{6C8EBF}
\definecolor{plum}{HTML}{ff7fff}
\definecolor{blueviolet}{HTML}{8a2be2}
\definecolor{lightgray}{rgb}{0.95, 0.95, 0.95}
\definecolor{amber}{rgb}{1.0, 0.75, 0.0}

\newcommand{\hgrow}{\rowcolor{lightgray!70}}

\usepackage{pifont}
\newcommand{\xmark}{\textcolor{red}{\ding{55}}}
\newcommand{\cmark}{\textcolor{darkgreen}{\ding{51}}}

\usepackage{arydshln}
\usepackage{siunitx}
\usepackage{ulem}

\usepackage{multirow} 
\usepackage{tikz}
\usetikzlibrary{tikzmark}
\makeatletter
\newcommand*\myfontsize{%
  \@setfontsize\myfontsize{6.7}{8}%
}
\makeatother

\definecolor{cadmiumgreen}{rgb}{0.0, 0.42, 0.24}

\definecolor{myred}{rgb}{0.7, 0.3, 0.0}
\definecolor{myblue}{rgb}{0.2, 0.3, 0.6}

\definecolor{baselinecolor}{gray}{.9}

\newcommand{\cmdrtaskname}{\texttt{CMDR}\xspace}
\newcommand{\cmdrdatasetname}{\texttt{CMDR-Bench}\xspace}
\newcommand{\cmdrmethodname}{\texttt{CMDR-Embed}\xspace}
\newcommand{\cmdrsynthname}{\texttt{CMDR-Synth}\xspace}
\newcommand{\cmdrembedpali}{\texttt{CMDR-Embed}$_{\text{Pali}}$\xspace}
\newcommand{\cmdrembedqwen}{\texttt{CMDR-Embed}$_{\text{Qwen}}$\xspace}
\newcommand{\cmdrnamelong}{\textbf{C}ontextual \textbf{M}ultimodal \textbf{D}ocument \textbf{R}etrieval\xspace}

\definecolor{grpA}{HTML}{E8F4FF}     
\definecolor{grpB}{HTML}{FFE8EB}     

\begin{document}

\title{CMDR:\\Contextual Multimodal Document Retrieval} 

\titlerunning{CMDR: Contextual Multimodal Document Retrieval}

\author{Ryota Tanaka\orcidlink{0009-0004-6619-0502} \and
Taku Hasegawa\orcidlink{0009-0003-2037-2717} \and
Kyosuke Nishida\orcidlink{0000-0002-8443-0651}}

\authorrunning{R.~Tanaka et al.}

\institute{Human Informatics Labs., NTT, Inc., Tokyo, Japan \\
\email{\{ryota.tanaka,taku.hasegawa,kyosuke.nishida\}@ntt.com}} 

\maketitle

\begin{abstract}
Multimodal document retrieval aims to retrieve relevant pages while preserving both textual and visual content from the original document. However, existing benchmarks primarily evaluate simple lexical or semantic matching, and most methods encode pages independently. Consequently, they overlook the contextual information in the document required to resolve queries that aggregate information across multiple pages. In this paper, we introduce \cmdrtaskname and \cmdrdatasetname, a new multimodal document retrieval task and benchmark that require modeling document context. To address this challenge, we propose \cmdrmethodname, a contextual multimodal embedding framework that explicitly incorporates document context by jointly encoding multiple pages and deriving page-level embeddings from a shared contextual representation. Furthermore, we introduce CMCL, a contextual multimodal contrastive learning objective for effectively training \cmdrmethodname, which balances contextual modeling with page-level discriminability. 
Experiments demonstrate that \cmdrmethodname significantly outperforms non-contextual embeddings, highlighting the importance of context-aware multimodal embeddings for advancing document retrieval\footnote{\faGlobe\ \textbf{Project Page:} \url{https://cmdr-bench.github.io}}.
\end{abstract}

\section{Introduction}
Real-world documents are inherently multimodal; they contain both textual and visual elements, with their content distributed across different regions. The ability to accurately retrieve information relevant to a user query from such diverse documents is crucial for various applications such as search engines~\cite{brin1998anatomy,liu2009learning} and Retrieval-Augmented Generation (RAG)~\cite{lewis2020retrieval,guu2020retrieval}. Unlike traditional text-based retrieval methods~\cite{robertson2009probabilistic,izacard2021unsupervised,lee2025nv}, which focus on textual data, multimodal document retrieval~\cite{SlideVQA2023,ma2024mmlongbench,faysse2024colpali,yu2024visrag,tanaka2025vdocrag} has emerged as a promising approach that seamlessly integrates multimodal information by treating documents as images.

\begin{figure*}[t!]
    \centering
    \includegraphics[width=0.99\textwidth]{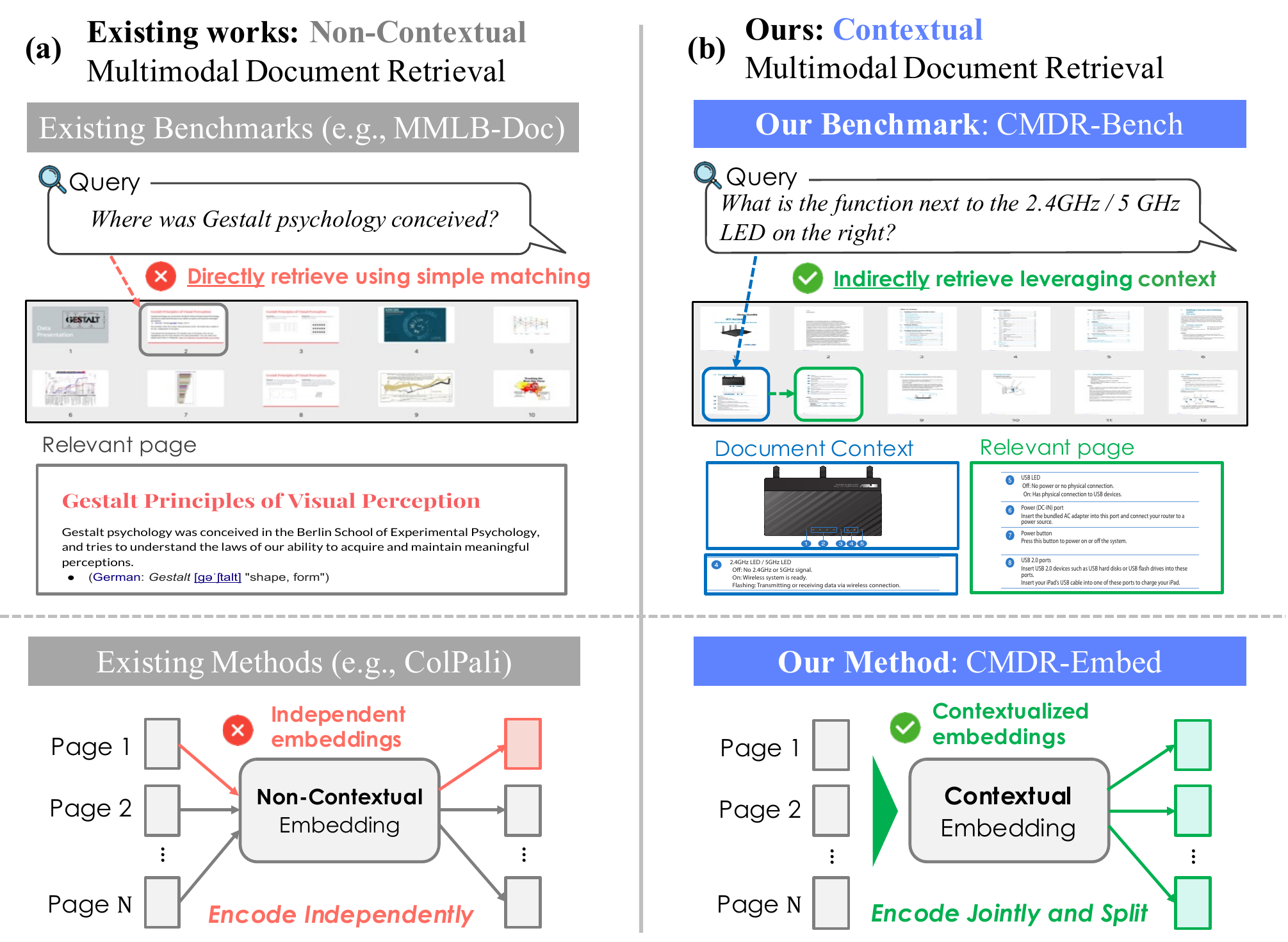}
    \caption{\textbf{Comparison benchmarks and methods}. Unlike previous benchmarks that rely on direct keyword or simple semantic matching, \cmdrdatasetname focuses on indirectly retrieving relevant pages by leveraging document context. In contrast to non-contextual embeddings that encode each page independently, \cmdrmethodname jointly encodes multiple pages to capture cross-page contextual relationships.}
    \label{fig:intro}
\end{figure*}

However, as illustrated in Figure~\ref{fig:intro}\textcolor{red}{a}, existing multimodal document retrieval benchmarks~\cite{SlideVQA2023,ma2024mmlongbench,deng2025longdocurl,chen2025visr} primarily evaluate direct retrieval using simple lexical or semantic matching between queries and individual pages. As a result, they do not assess retrieval capabilities that leverage \textit{document context}, such as cross-page dependencies and document global structure. Moreover, existing retrieval methods~\cite{faysse2024colpali,yu2024visrag,ma2024unifying,liu2025any,tanaka2025vdocrag} independently encode each page in multi-page documents, implicitly assuming that contextualization yields negligible benefit. In practice, however, important information is often distributed across pages, requiring models to reason over multiple pages to fully resolve a query.

Motivated by these limitations, we propose a new multimodal document retrieval task, \cmdrnamelong (\cmdrtaskname), which aims to indirectly retrieve relevant pages from a multimodal document by modeling context across multiple pages (see Figure~\ref{fig:intro}\textcolor{red}{b}). To facilitate research in this area, we introduce a new benchmark, \cmdrdatasetname, comprising 800 high-quality queries carefully annotated by humans across four categories. The documents span six domains, with an average length of 183.5 pages. \cmdrdatasetname is the first dataset that jointly requires multi-page reasoning and multimodal document retrieval.

To address this task, we introduce \cmdrmethodname, a novel embedding model that captures contextual information by jointly encoding multiple pages and then splitting the representations at the page level. While joint encoding enables context modeling, it also causes information leakage across pages within the same document, leading to reduced page-level discriminability. To address this issue,  \cmdrmethodname is trained with a new contrastive learning framework, \textbf{C}ontextual \textbf{M}ultimodal \textbf{C}ontrastive \textbf{L}earning (CMCL), which balances effective use of contextual information and page discriminability by incorporating context-aware hard negatives sampled from the same document. Experiments demonstrate that \cmdrmethodname significantly outperforms non-contextual embedding models, with minimal computational overhead. This highlights the importance of context-aware multimodal embeddings for advancing document retrieval. 

Our main contributions are summarized as follows:
\begin{itemize}
    \item \cmdrtaskname and \cmdrdatasetname, a new multimodal document retrieval task and benchmark that require retrieving relevant pages from a long multi-page document by modeling contextual information across multiple pages.
    \item \cmdrmethodname, a new multimodal document embedding model that jointly encodes multiple pages and then produces page-level representations to effectively capture document context.
    \item CMCL, a new contextual multimodal contrastive learning framework for training \cmdrmethodname, which leverages context-aware hard negatives to balance contextual understanding and page discriminability.

\end{itemize}
\section{Related Work}
\medskip\noindent\textbf{Benchmarks for multimodal document retrieval.} 
The growing interest in Document RAG~\cite{yu2024visrag,ma2025visa,tanaka2025vdocrag,chen2025document,dong2025benchmarking,tong2025hkrag,zheng2025retrieval,yan2026docseeker,wang2026vrag} beyond single-page DocumentVQA~\cite{Mathew_2021_WACV,DBLP:conf/aaai/TanakaNY21,Mathew_2022_WACV,zhu2022towards,ChenZCJZLX021} necessitates comprehensive benchmarks for multimodal document retrieval.
Existing benchmarks~\cite{tito2023hierarchical,SlideVQA2023,dong2025mmdocir,faysse2024colpali,mace2025vidore,tanaka2025vdocrag,zhang2025ocr,chen2025visr} predominantly focus on simple textual or semantic matching between queries and individual pages. As a result, they fail to evaluate retrieval capabilities that depend on document context, such as information spanning multiple pages or document-level structures.
Although multi-hop retrieval benchmarks~\cite{SlideVQA2023,ma2024mmlongbench,deng2025longdocurl} have been explored, they are largely limited to direct retrieval based on explicit query–page matches (e.g., ``What proteins are involved in pathway X, and what experiments validate their roles?''). In contrast, our benchmark uniquely targets indirect multimodal retrieval, requiring models to identify relevant pages that are not explicitly mentioned in the query. Further comparisons are provided in Section~\ref{sec:comparison_datasets}.

\medskip\noindent\textbf{Models for multimodal document retrieval.}
There is a growing trend toward building Large Vision–Language Models (LVLMs)~\cite{alayrac2022flamingo,abdin2024phi,achiam2023gpt,bai2025qwen2} that integrate image understanding capabilities by combining an LLM and an image encoder~\cite{radford2021learning,li2022blip,zhai2023sigmoid}.
Multimodal document retrieval methods~\cite{faysse2024colpali,ma2024unifying,tanaka2025vdocrag,yu2024visrag,teiletche2025modernvbert} based on LVLMs directly encode document images, thereby preserving both textual content and layout information while bypassing OCR.
Despite advances, existing multimodal document retrievers still struggle to model cross-page relationships, as they typically encode each page independently, without explicitly capturing document context.
In contrast, our method addresses this limitation by jointly encoding multiple pages and subsequently splitting them to incorporate cross-page context, and learning context-aware embeddings while preserving discriminability among pages within the same document.

\medskip\noindent\textbf{Contextual retrieval.} 
Contextual retrieval is crucial for accurately retrieving information that depends on the surrounding context rather than isolated passages, especially in long-context documents. To evaluate such capabilities, context-aware passage retrieval benchmarks have been proposed~\cite{wang2024dapr,conti-etal-2025-context}. In parallel with benchmark development, a few studies have explored context-aware indexing methods using late chunking~\cite{gunther2024late,conti-etal-2025-context} to incorporate contextual information. Although our method is also inspired by late chunking, it differs in that we introduce a novel contrastive learning objective that jointly optimizes context-aware embeddings while preserving page-level discriminability. In addition, existing benchmarks and methods mainly focus on retrieving passages from plain-text documents. In contrast, we address the challenge of retrieving pages from visually-rich documents organized in complex, multimodal formats.

\section{Our Benchmark: \cmdrdatasetname}
\subsection{Task Definition of \cmdrtaskname}
We present \cmdrnamelong (\cmdrtaskname), a new multimodal document retrieval task in which a system must retrieve relevant pages from a multi-page document (i.e., an ordered set of document images) while leveraging the document context. Formally, given a query $q$ and a multi-page document $\mathcal{D} = \{I_1, \dots, I_n\}$, the task is to retrieve top $k$ pages $\mathcal{D}^+ \subset \mathcal{D}$ ($k \ll n$) containing the answer, where identifying $\mathcal{D}^+$ requires reasoning over the document context rather than relying on the contents of $\mathcal{D}^+$ alone. The context pages are the other pages that provide the necessary information to identify the relevant pages, but do not contain the answer. 
Note that even when the answer spans multiple pages, the task is ``direct'' retrieval and falls outside the scope of \cmdrtaskname if the query directly and explicitly references the content of those pages (e.g., \textit{``What are the items of A and B?''}).

\begin{figure*}[t!]
    \centering
    \includegraphics[width=0.92\textwidth]{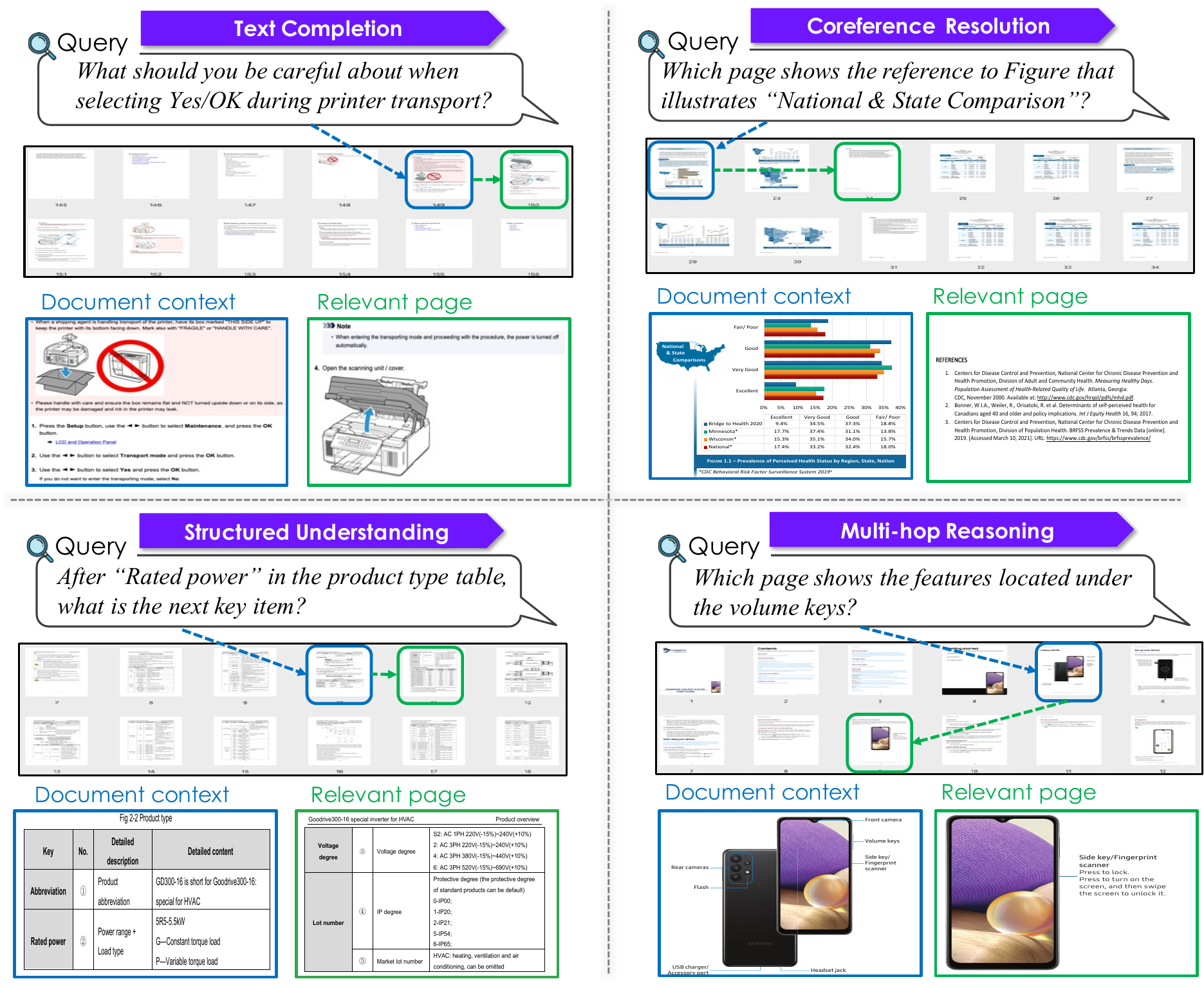}
    \caption{\textbf{Dataset examples in \cmdrdatasetname}. \cmdrdatasetname comprises four subtasks that focus on retrieving pages from a multi-page document using document context.}
    \label{fig:examples}
\end{figure*}

\medskip\noindent\textbf{Importance of context.}
As illustrated in Figure~\ref{fig:examples}, document context plays a critical role in retrieval by resolving ambiguities and identifying relevant content that cannot be inferred from a single page. In real-world documents, relevant information is often distributed across multiple pages, with key entities, conditions, or visual elements introduced on one page and clarified on others.

\subsection{Data Collection of \cmdrdatasetname} 

\medskip\noindent\textbf{Document collection.}
\label{para:document_lategory}
As a retrieval benchmark, the dataset should include documents that cover diverse domains and contain sufficiently long page sequences. To this end, we collected a variety of PDF documents from Common Crawl~\cite{pdfa} and Manualslib~\cite{manualslib}, filtering them by page length (>100 pages) and language (English). Following prior work~\cite{ma2024mmlongbench,deng2025longdocurl}, we manually categorized the collected documents into six domains: \textit{manuals}, \textit{research reports \& papers}, \textit{theses \& dissertations}, \textit{project proposals}, \textit{books \& e-books}, and \textit{meeting minutes \& summaries}. Each document is then rendered page-by-page as PNG images.

\medskip\noindent\textbf{Query annotation.}
\label{para:query_lategory}
As illustrated in Figure~\ref{fig:examples}, prior to query annotation, we predefined four query categories to assess the ability to retrieve pages using document context as follows: 
\begin{description}
\item[(1) Text Completion (TC):] Key textual content is split across pages, requiring accurate reconstruction of the semantic flow to maintain coherence.
\item[(2) Coreference Resolution (CR):] References such as pronouns or abbreviations must be resolved using contextual cues found only in other pages.
\item[(3) Structured Understanding (SU):] Understanding how structured components relate to each other across page boundaries (e.g., a table header defined on one page and its rows continued on the next), independent of entity resolution or multi-entity aggregation.
\item[(4) Multi-hop Reasoning (MR):] An inference path that connects the entities in the query and the relevant page via the document context. Unlike CR, which resolves referential links between an entity and its mentions, MR involves multi-entity semantic reasoning. 
\end{description}

To ensure query quality, the authors, who hold PhDs in computer science and are proficient in English, first carefully read each page to understand its structure and cross-page dependencies. Then, they selected relevant pages and created at least three queries per document, each explicitly requiring contextual reasoning. The query categories are conceptually distinct, and each query is assigned to a primary category to avoid overlap in their reasoning requirements.

\medskip\noindent\textbf{Data quality control and refinement.}
To ensure that queries genuinely evaluate contextual multimodal retrieval, we used state-of-the-art retrievers solely as a \textit{supporting tool} to help human annotators identify trivial or erroneous annotations. Specifically, we retrieved top-ranked pages from a document using three non-contextual retrievers~\cite{lee2025nv,jiang2024vlm2vec,faysse2024colpali}, and computed the average Recall@1 score for each query. Queries with a score of 1.0 were frequently identified (via manual inspection) as trivial, meaning they could be resolved without leveraging document context. In such cases, annotators were asked to revise the queries to incorporate stronger cross-page dependencies. Conversely, queries with low scores were carefully reviewed to distinguish truly difficult contextual cases from those with annotation mistakes (e.g., wrong gold page or ambiguous phrasing). Erroneous cases were corrected or discarded. During this refinement process, 16\% of queries were removed, and 4\% were revised. This process ensures that the final benchmark emphasizes non-trivial contextual reasoning, while keeping human judgment as the primary authority throughout annotation.

\begin{figure}[t]
    \begin{minipage}{0.49\textwidth}
        \begin{minipage}{\linewidth}
            \centering
            \begin{subfigure}{0.49\linewidth}
                \includegraphics[width=\linewidth, trim={0.7em 0 0.6em 0em}, clip]{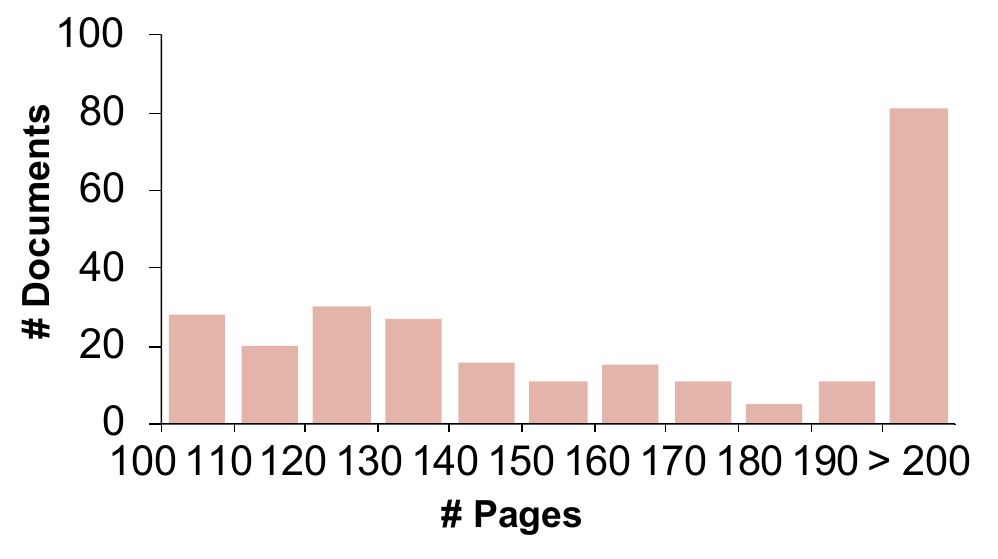}
                \caption{\#Pages Per Doc.}
                \label{fig_dist:subimg1}
            \end{subfigure}
            \begin{subfigure}{0.49\linewidth}
                \includegraphics[width=\linewidth, trim={0.7em 0 0.6em 0em}, clip]{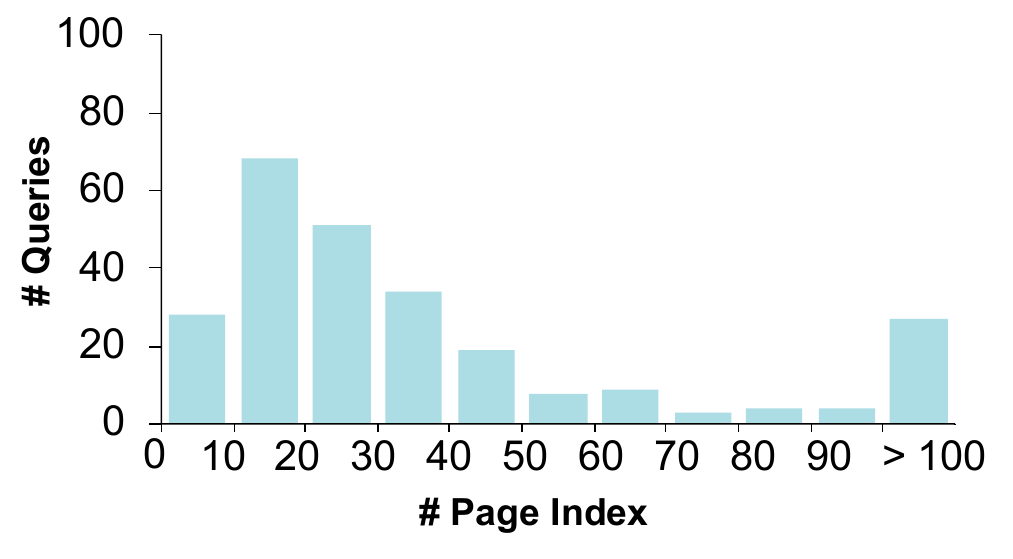}
                \caption{Related Page Index.}
                \label{fig_dist:subimg2}
            \end{subfigure}
            \begin{subfigure}{0.48\linewidth}
                \includegraphics[width=\linewidth, trim={0.7em 0 0.6em 0em}, clip]{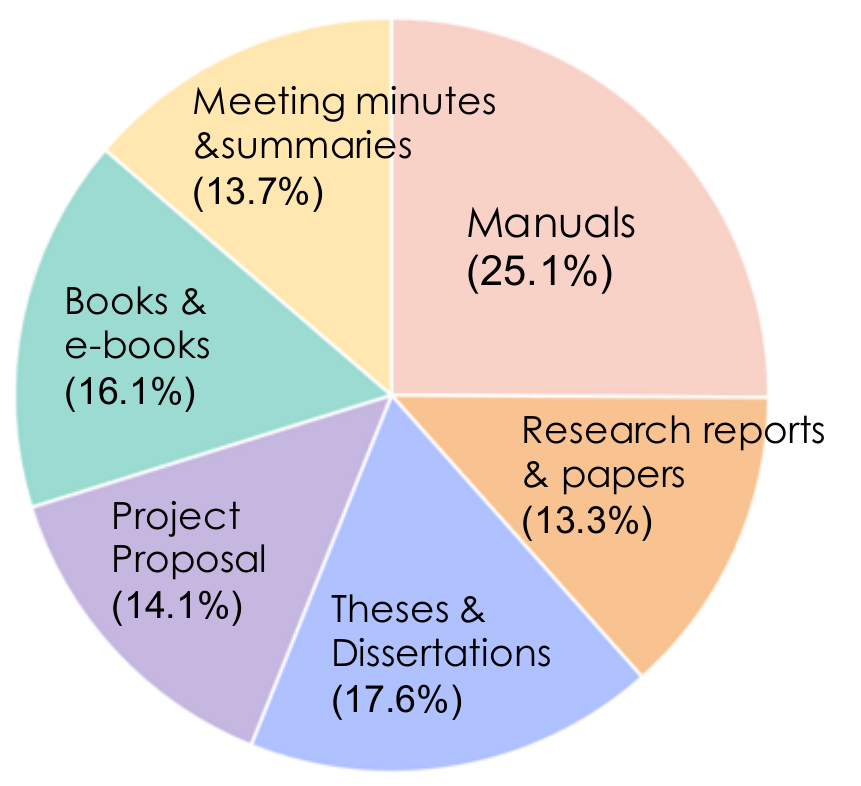}
                \caption{Doc. Categories.}
                \label{fig_dist:subimg3}
            \end{subfigure}
            \begin{subfigure}{0.45\linewidth}
                \includegraphics[width=\linewidth, trim={0.7em 0 0.6em 0em}, clip]{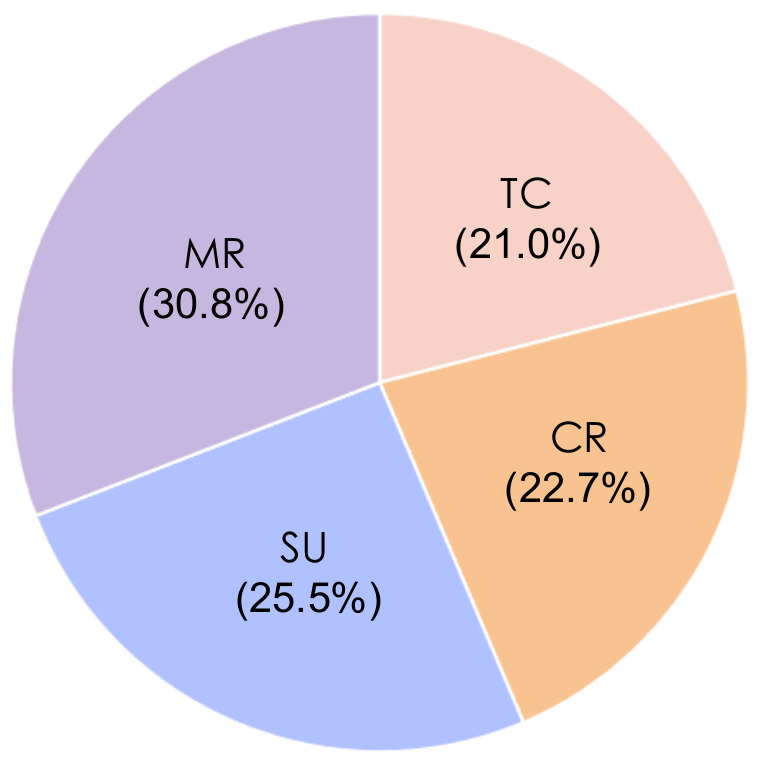}
                \caption{Query Categories.}
                \label{fig_dist:subimg4}
            \end{subfigure}
            \captionof{figure}{\textbf{Dataset distribution}.}
            \label{fig:distributions}
        \end{minipage}
    \end{minipage}
 \begin{minipage}{0.47\textwidth}
    \centering
    \captionof{table}{\textbf{Comparison benchmarks}. Query annotation: M (Manual), S (Synthesized), R (Repurposed).}
        \scalebox{0.7}{
    \tabcolsep=1.0pt
    \small
    \begin{tabular}{lccccccc} 
        \toprule
         \multirow{2}{*}{Benchmarks} & Require & Multi & Multi-hop & Query & \#Avg. \\ & Context & Modal & Type & Annot. & Pages\\ \midrule
        DAPR~\cite{wang2024dapr} & \cmark & \xmark & Indirect & R & 1.0 \\
        ConTEB~\cite{conti-etal-2025-context} & \cmark & \xmark & -- & R & 1.0 \\
        MP-DocVQA~\cite{tito2023hierarchical} & \xmark & \cmark & -- & R & 8.1 \\
        SlideVQA~\cite{SlideVQA2023} & \xmark & \cmark & Direct &  M+R  &  20.0 & \\
        MMLB-Doc~\cite{ma2024mmlongbench} & \xmark & \cmark & Direct & M+R  & 47.5 \\
        LongDocURL~\cite{deng2025longdocurl} & \xmark & \cmark & Direct & S  & 85.6 \\
        MMDocIR~\cite{dong2025mmdocir} & \xmark & \cmark & Direct & R  & 65.1 \\
        ViDoRe~\cite{faysse2024colpali} & \xmark & \cmark & -- & S+R  & 1.0 \\
        ViDoRe v2~\cite{mace2025vidore} & \xmark & \cmark & -- & S+R  & 1.0 \\
        OpenDocVQA~\cite{tanaka2025vdocrag} & \xmark & \cmark & Direct & S+R  & 3.1 \\
        OHR-Bench~\cite{zhang2025ocr} & \xmark & \cmark & -- & S & 6.8 \\
        VisR-Bench~\cite{chen2025visr} & \xmark & \cmark & -- & S & 4.5 \\ \midrule
        \cmdrdatasetname (Ours) & \cmark & \cmark & Indirect & M & 183.5 \\
        \bottomrule
    \end{tabular}
    }
    \label{tab:comparison}
\end{minipage}
\end{figure}

\subsection{Dataset Statistics and Comparison with Related Datasets}
\label{sec:comparison_datasets}
In total, \cmdrtaskname contains 800 queries across four categories, with 255 documents spanning six different types. Detailed distributions are shown in Figure~\ref{fig:distributions}. 

Table~\ref{tab:comparison} compares \cmdrdatasetname with existing related benchmarks. \cmdrdatasetname offers three unique key properties: (1) \textbf{\cmdrdatasetname is the first benchmark that jointly requires contextual reasoning and multimodal document retrieval}. While DAPR~\cite{wang2024dapr} and ConTEB~\cite{conti-etal-2025-context} involve contextual reasoning, they are limited to single-page and text-only retrieval settings. Additionally, existing multimodal document retrieval datasets~\cite{tito2023hierarchical,SlideVQA2023,ma2024mmlongbench,dong2025mmdocir,tanaka2025vdocrag,zhang2025ocr,chen2025visr} handle multi-page documents but mainly focus on \textit{direct} retrieval from explicit query-page matches. \cmdrdatasetname uniquely targets \textit{indirect} cross-modal retrieval, which requires identifying relevant pages that are not explicitly mentioned in the query. (2) \textbf{All queries in \cmdrdatasetname are manually annotated, ensuring high-quality and better reflect real-world scenarios.} In contrast, several existing datasets rely mainly on synthesized (e.g., VisR-Bench~\cite{chen2025visr}) or repurposed annotations (e.g., DAPR~\cite{wang2024dapr}), which may not fully align with retrieval-specific task objectives.  (3) \textbf{The documents in \cmdrdatasetname are significantly longer, offering a challenging retrieval task.} The average page length reaches 183.5 pages, which is over 2.1$\times$ that of LongDocURL~\cite{deng2025longdocurl} (85.6 pages).

\section{Our Model: \cmdrmethodname}

\begin{figure*}[t!]
    \centering
\includegraphics[width=.99\textwidth]{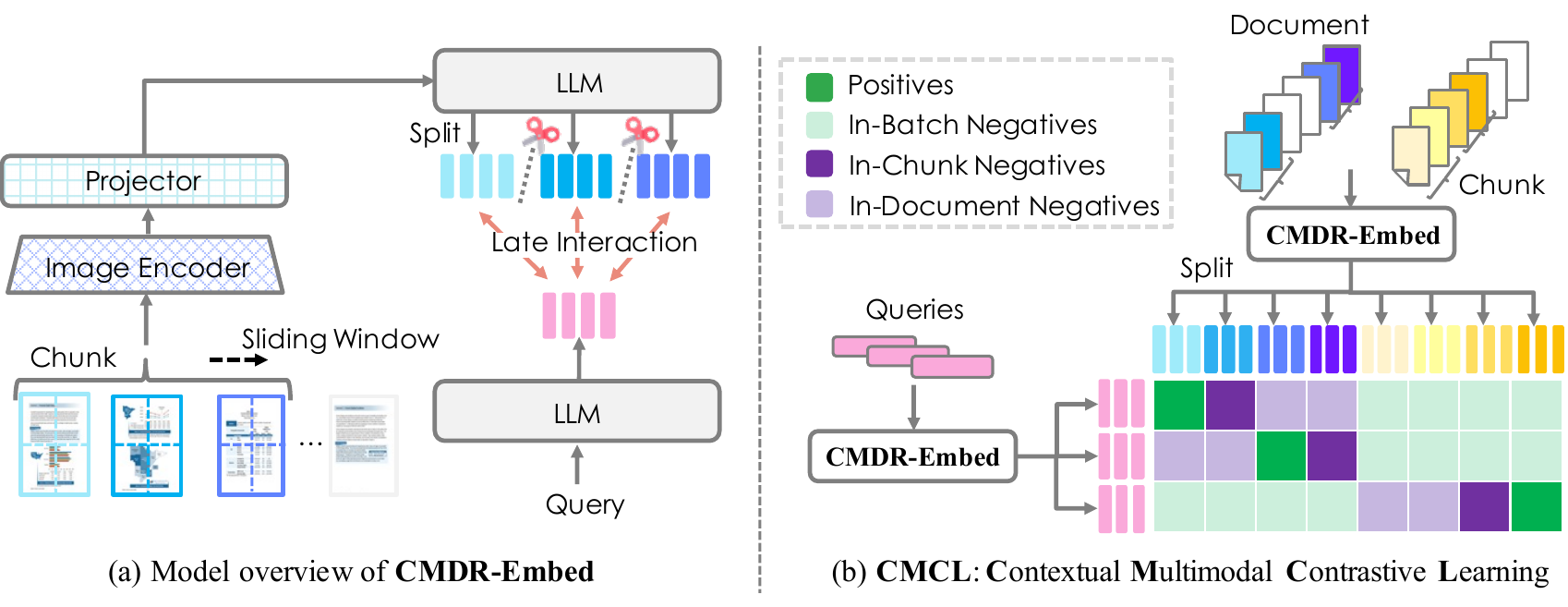}
    \caption{\textbf{Our model and training loss}. (a) \cmdrmethodname encodes multiple pages jointly and then splits them to capture contextual information. (b) CMCL loss optimizes context-aware embeddings while preserving discriminability within the same document.}
    \label{fig:cmde}
\end{figure*}

\subsection{Architecture Overview}
Figure~\ref{fig:cmde}\textcolor{red}{a} shows an overview of our model, called \cmdrmethodname.
Unlike non-contextual retrievers like ColPali~\cite{faysse2024colpali}, which encode each page independently and thus fail to capture cross-page context, \cmdrmethodname encodes not only the content of the page itself but also explicitly incorporates contextual information.

\medskip\noindent\textbf{Contextualized multimodal encoding.} 
To enable cross-page interactions within a document, we adopt a \textit{chunk-then-split} strategy. Specifically, consecutive pages are first encoded together, and the resulting representations are then split into page-level embeddings (i.e., sequences of token-level embeddings). Formally, given a document with $N$ pages $\{I_1, \ldots, I_N\}$, we divide it into multiple chunks using a sliding window with a stride length $s$ and a window size $w$. The $t$-th chunk is defined as $\{I_{s(t-1)+1}, \ldots, I_{s(t-1)+w}\}$. Each chunk is fed into an LVLM to produce token-level representations, which are then separated into page-level embeddings $\mathbf{E}^{\mathbf{I}}_n \in \mathbb{R}^{N^I \times D}$, where $D$ denotes the embedding dimension and $N^I$ represents the number of token embeddings for page $I_n$. When chunks overlap (i.e., $s < w$), tokens may appear in multiple chunks. In such cases, we use the representations from the earlier chunk to avoid redundant computation. Finally, we obtain a sequence of contextualized page embeddings $\{\mathbf{E}^{\mathbf{I}}_1, \ldots, \mathbf{E}^{\mathbf{I}}_N\}$.

\medskip\noindent\textbf{Multimodal late interaction.}
Following prior work~\cite{faysse2024colpali,teiletche2025modernvbert}, we compute similarities between the query and each page using Late Interaction (LI)~\cite{khattab2020colbert,chen2024bge}, which captures fine-grained query–page interactions via multi-vector matching. 
Let the query vectors be represented by $\mathbf{E}^\mathbf{q} \in \mathbb{R}^{N^q \times D}$, where $N^q$ denotes the number of query tokens. The LI score $\mathbf{LI}(q, I_n)$ is the sum over all query vectors $\mathbf{E}^{\mathbf{q}_i}$, of its maximum dot product $\langle\cdot,\cdot\rangle$ with the contextualized page vectors $\mathbf{E}^{\mathbf{I}_j}_n$. 

\begin{equation}
  \mathbf{LI}(q, I_n) = \sum_{i \in [1, N^q]} \max_{j\in [1, N^I]} \langle\mathbf{E}^{\mathbf{q}_i},\mathbf{E}^{\mathbf{I}_j}_n\rangle. 
\end{equation}


\subsection{Contextual Multimodal Contrastive Learning} 
Because \cmdrmethodname encodes multiple pages jointly, representations from different pages in the same document can be mixed, which weakens page-level discriminability. To address this issue, we propose a new contrastive learning framework, \textbf{C}ontextual \textbf{M}ultimodal \textbf{C}ontrastive \textbf{L}earning (CMCL). Specifically, as shown in Figure~\ref{fig:cmde}\textcolor{red}{b}, we reinforce the standard InfoNCE objective~\cite{oord2018representation} by introducing two types of context-aware hard negatives: \textit{In-Chunk Negatives} $\mathcal{I}_{\text{chunk}}$, which are different pages within the same chunk, and \textit{In-Document Negatives} $\mathcal{I}_{\text{doc}}$, which are other chunks from the same document. Intuitively, $\mathcal{I}_{\text{chunk}}$ is more effective when similar context appears on nearby pages, whereas $\mathcal{I}_{\text{doc}}$ is particularly helpful when information similar to the relevant page appears on distant pages outside the current chunk. This design prevents neighboring pages from becoming overly similar and ensures that each page maintains a distinct representation, even when contextual information is shared.


\medskip\noindent\textbf{CMCL loss.}
By jointly leveraging these two types of hard negatives, the CMCL loss is defined as: $\mathcal{L}_\text{CMCL} = \lambda \mathcal{L}_{\text{Context}} + (1- \lambda )\mathcal{L}_{\text{Batch}}$, where $\lambda$ balances the two loss terms. Let $I^{+}$ denote the positive page corresponding to a query $q$, and let $I^-_n$ denote a sampled negative page. Each loss term is computed as follows:
\begin{equation}
    \mathcal{L}_{\text{Context}} = -\text{log} \frac{\exp(\mathbf{LI}(q, I^+)/\tau)} {\exp(\mathbf{LI}(q, I^+)/\tau) + \sum_{I^-_n\in \mathcal{I_{\text{chunk}}}\cup\mathcal{I_{\text{doc}}}} \exp(\mathbf{LI}(q, I^-_n)/\tau)}, 
\end{equation}
\begin{equation}
    \mathcal{L}_{\text{Batch}} = -\text{log} \frac{\exp(\mathbf{LI}(q, I^+)/\tau)} {\exp(\mathbf{LI}(q, I^+)/\tau) + \sum_{I^-_n\in \mathcal{I_{\text{batch}}}} \exp(\mathbf{LI}(q, I^-_n)/\tau)},
\end{equation}
where $\tau$ is a hyperparameter that scales the late-interaction score, and $\mathcal{I}_{\text{batch}}$ denotes the set of in-batch negatives that do not belong to $\mathcal{I}_{\text{chunk}}$ and $\mathcal{I}_{\text{doc}}$.

\section{Experiments}

\subsection{Experimental Setup}

\medskip\noindent\textbf{Training dataset.} 
\label{para:training_dataset}
We constructed the \cmdrsynthname dataset, which provides 39,796 query-page pairs for training contextual multimodal retrieval models. 
For PDFs collected from Common Crawl that are not included in \cmdrdatasetname, we created queries through a four-stage process. 
(1) Using Qwen2.5VL 72B~\cite{bai2025qwen2}, we classified each page based on whether its content requires contextual information, i.e., whether the page is self-contained. 
(2) Pages identified as requiring context were then used as input to UniSE~\cite{liu2025any} to retrieve pages that provide relevant contextual information. This step constructs page pairs consisting of a contextual page and a relevant page. 
(3) The constructed page pairs are fed into Qwen2.5VL 72B to generate queries requiring contextual information for retrieval. 
(4) To ensure the quality of the synthesized data, we performed filtering using Qwen2.5VL 72B to assess query-page relevance. Detailed information is provided in the Supplementary Material.

\medskip\noindent\textbf{Implementation details.} We initialized \cmdrmethodname with the finetuned weights of ColPali~\cite{faysse2024colpali} or ColQwen~\cite{faysse2024colpali}, a state-of-the-art vision-based retrieval model. We used LoRA~\cite{hu2021lora} with $\alpha$ = 32 and $r$ = 32 for the LLM layers and the final projection layer. We trained \cmdrmethodname for three epochs on eight A100-80G GPUs with AdamW~\cite{loshchilov2017decoupled} optimizer and FlashAttention~\cite{dao2022flashattention}, using batch sizes of 192 and a learning rate of 2e-4. We set the temperature $\tau$ to 0.02 and chose $\lambda$ to 0.5 from $\{0.1, 0.5, 0.9\}$ for CMCL loss. For chunks, we set $s=2$ and $w=4$ during both training and inference. We measured indexing latency on a single A100-80G GPU and search latency on an AMD EPYC 7742 64-core CPU.

\medskip\noindent\textbf{Baselines.}
We compare \cmdrmethodname with three categories of non-contextual retrievers that encode pages independently, without leveraging document context.
(1) \textit{Text Retrievers} encode documents using the OCR text extracted from Tesseract~\cite{smith2007overview}. This category includes \textbf{BM25}~\cite{lu2024bm25s}, a lexical matching model; \textbf{Contriever}~\cite{izacard2021unsupervised}, \textbf{BGE}~\cite{xiao2024c}, and \textbf{E5}~\cite{wang2022text}, which are BERT-based text embedding models~\cite{DevlinCLT19}; and \textbf{NV-Embed-v2}~\cite{lee2025nv}, a state-of-the-art LLM-based embedding model.
(2) \textit{General Multimodal Retrievers} are trained on diverse multimodal retrieval tasks, such as image retrieval and visual question answering. This category includes \textbf{CLIP}~\cite{radford2021learning} and \textbf{SigLIP}~\cite{zhai2023sigmoid}, which are dual-encoder vision–language models, as well as \textbf{E5-V}~\cite{jiang2024e5}, \textbf{VLM2Vec}~\cite{jiang2024vlm2vec}, \textbf{GME}~\cite{Zhang_2025_CVPR}, and \textbf{Qwen3-VL Embedding}~\cite{li2026qwen3}, which are LVLM-based multimodal embedding models.
(3) \textit{Multimodal Document Retrievers} are built upon LVLMs and specifically trained for document retrieval. \textbf{ColModernVBERT}~\cite{teiletche2025modernvbert}, \textbf{DSE}~\cite{ma2024unifying}, \textbf{VisRAG-Ret}~\cite{yu2024visrag}, and \textbf{UniSE}~\cite{liu2025any} are single-vector embeddings, whereas \textbf{ColPali}~\cite{faysse2024colpali} and \textbf{ColQwen}~\cite{faysse2024colpali} are multi-vector representations.
To verify the effectiveness of leveraging contextual information, we finetune our backbones on the \cmdrsynthname dataset (\textbf{ColPali+Finetune}, \textbf{ColQwen+Finetune}).

\medskip\noindent\textbf{Evaluation metrics.}
We evaluate performance using \textbf{nDCG@5}, a widely used metric in retrieval~\cite{kamalloo2023resources,faysse2024colpali}. We also report overall performance, which computes the average scores across four query categories defined in Section~\ref{para:query_lategory}.

\begin{table*}[t!]
    \centering
    \caption{\textbf{Contextual retrieval results} (nDCG@5) on \cmdrdatasetname. The performance gain in \textcolor{darkgreen}{green} is compared to the same finetuned backbone. Overall performance reports the average scores across four query categories defined in Section~\ref{para:query_lategory}.} 
        \scalebox{0.81}{
    \tabcolsep=2.5pt
    \small
    \begin{tabular}{llllllll} 
        \toprule
        \multirow{2}{*}{Model} & \multirow{2}{*}{Backbone} & \multirow{2}{*}{\#Params} & \multicolumn{4}{c}{Query Categories} & \multirow{2}{*}{Overall} \\ 
        & & & TC & CR & SU & MR  \\
        \midrule
        \multicolumn{2}{>{\columncolor{grpB}}c}{\textbf{Non-Contextual Models}} & \multicolumn{2}{c}{\textit{Text Retrievers}} \\ 
        BM25~\cite{lu2024bm25s} & -- & -- & 29.1 & 19.5 & 20.9 & 24.9 & 23.6 \\
        Contriever~\cite{izacard2021unsupervised} & BERT-base &  109M & 35.7 & 15.0 & 23.5  & 26.4 & 25.1 \\
        BGE~\cite{xiao2024c} & BERT-base & 109M  & 42.8  & 18.4  & 28.5  & 29.5 & 29.8 \\
        E5~\cite{wang2022text} & BERT-large & 340M & 36.4 & 19.0 & 26.5 & 27.1 & 27.2 \\
        NV-Embed-v2~\cite{lee2025nv} & Mistral-7B &  7.9B  & 36.2 & 16.9 & 28.6 & 30.4 &  28.0 \\ \hdashline
        \multicolumn{8}{c}{\textit{General Multimodal Retrievers}} \\ 
        CLIP~\cite{radford2021learning} & CLIP-large & 428M & 13.3 & 5.8 & 10.6 & 13.2 & 10.7 \\
        SigLIP~\cite{zhai2023sigmoid} & SOViT-400m & 878M & 24.5 & 8.7 & 15.5 & 22.2 & 17.8 \\
        E5-V~\cite{jiang2024e5} & LLaVA-1.6 & 8.4B  & 35.4  & 22.8  & 31.2  & 34.8  & 31.0  \\
        VLM2Vec~\cite{jiang2024vlm2vec} & Phi-3.5-V & 4.2B & 27.9  & 16.4 & 26.4  & 28.2 & 24.7 \\ 
        GME~\cite{Zhang_2025_CVPR} & Qwen2-VL & 2.2B & 37.1  & 26.0 & 30.8& 38.0 & 33.0 \\
        Qwen3-VL Embedding~\cite{li2026qwen3} & Qwen3-VL & 8B & 38.2 & 26.6 & 35.7 & 37.2 & 34.4 \\
        \hdashline
        \multicolumn{8}{c}{\textit{Multimodal Document Retrievers}} \\ 
        ColModernVBERT~\cite{teiletche2025modernvbert} & ModernBERT & 250M  & 31.3 & 22.7  & 25.0  & 27.5 &  26.6 \\
        DSE~\cite{ma2024unifying} & Phi-3-V & 4.2B  & 33.7 & 23.0 & 29.1 & 30.4 & 29.1\\
        VisRAG-Ret~\cite{yu2024visrag} & MiniCPM-V & 3.4B  & 35.3  & 19.8 & 27.8  & 33.9 & 29.2 \\
        UniSE~\cite{liu2025any} & Qwen2-VL & 2.2B &  32.3 & 24.6  & 30.9 &33.0  & 30.2 \\
        ColPali~\cite{faysse2024colpali} & Paligemma & 2.9B  & 39.1  & 24.2 & 30.2 & 35.4 &  32.2 \\
        +Finetuned on \cmdrsynthname & Paligemma & 2.9B  & 41.0 & 27.5  & 36.8 & 39.9 & 36.3\\ 
        ColQwen~\cite{faysse2024colpali} & Qwen2-VL & 2.2B  & 38.0  & 28.9  & 30.0  & 35.9 & 33.2 \\ 
        +Finetuned on \cmdrsynthname & Qwen2-VL & 2.2B  & 45.8 & 34.8  & 38.7  & 42.2  & 40.4 \\
       \multicolumn{2}{>{\columncolor{grpA}}c}{\textbf{Contextual Models  (Ours)}} \\ 
        \cmdrembedpali & ColPali & 2.9B & 53.8$_{\textcolor{darkgreen}{\uparrow12.8}}$ & 33.3$_{\textcolor{darkgreen}{\uparrow5.8}}$ & 57.7$_{\textcolor{darkgreen}{\uparrow20.9}}$ & 54.3$_{\textcolor{darkgreen}{\uparrow14.4}}$ & 49.8$_{\textcolor{darkgreen}{\uparrow13.5}}$ \\
        \cmdrembedqwen & ColQwen & 2.2B & \textbf{64.6}$_{\textcolor{darkgreen}{\uparrow18.8}}$ & \textbf{38.4}$_{\textcolor{darkgreen}{\uparrow3.6}}$ & \textbf{64.7}$_{\textcolor{darkgreen}{\uparrow26.0}}$ & \textbf{58.9}$_{\textcolor{darkgreen}{\uparrow16.7}}$ & \textbf{56.6}$_{\textcolor{darkgreen}{\uparrow16.2}}$ \\
        \bottomrule
    \end{tabular}
    }
    \label{tab:retrieval_results}
\end{table*}

\subsection{Main Results}
\medskip\noindent\textbf{Do our contextual models outperform non-contextual models?}
As shown in Table~\ref{tab:retrieval_results}, \cmdrmethodname, which leverages contextual information, significantly outperforms models that do not consider the document context. Importantly, this performance gain cannot be attributed solely to the training data. Even when trained on the same data (\cmdrsynthname), the non-contextual models ColPali and ColQwen still underperform compared to our contextual models \cmdrembedpali and \cmdrembedqwen. \cmdrembedqwen emerges as the strongest retriever in this setting, achieving an average improvement of 16.2 points over the best non-contextual baseline. These results demonstrate that explicitly modeling document context yields benefits beyond what existing approaches can capture.

\medskip\noindent\textbf{What are the characteristics of \cmdrdatasetname?}
\label{rq:query_category}
Table~\ref{tab:retrieval_results} shows that multimodal document retrievers achieve strong performance across all categories. This demonstrates that \cmdrdatasetname requires methods to jointly understand the text, layout, and visual modalities of documents. Even in the text-related category (TC), text retrievers underperform vision-based retrievers. This is because many pages contain visual elements, such as charts, tables, and figures, that OCR cannot reliably parse. Errors in OCR can introduce noise into the textual representation, degrading retrieval performance. Moreover, all models generally perform worse on the CR and MR categories. This highlights that retrieval scenarios that require modeling long-range context, such as resolving references and aggregating information across multiple pages, remain particularly challenging for current multimodal retrievers. Overall, these findings indicate the importance of context-aware retrieval for complex, long-form, multi-page documents.

\begin{figure}[t]
\centering
\begin{minipage}[t]{0.5\textwidth}
    \vspace{0pt} 
    \centering
\captionof{table}{\textbf{Ablation studies of our key components}, including context-aware hard negatives, CMCL, and LI.}
\scalebox{0.78}{
\setlength{\tabcolsep}{3pt}
\small
\begin{tabular}{ll}
    Key Component & Overall  \\ \toprule
    \hgrow \cmdrembedpali & \textbf{49.8} \\ 
    w/o In-Chunk Negatives & 48.7$_{\textcolor{red}{\downarrow1.1}}$ \\ 
    w/o In-Document Negatives & 48.6$_{\textcolor{red}{\downarrow1.2}}$ \\ 
    w/o CMCL Loss ($\hookrightarrow$ only $L_{\text{Batch}}$) & 45.2$_{\textcolor{red}{\downarrow4.6}}$ \\
    w/o LI ($\hookrightarrow$ Mean Pooling+Cosine Sim.) & 23.3$_{\textcolor{red}{\downarrow26.5}}$ \\ \midrule
    \hgrow \cmdrembedqwen & \textbf{56.6} \\ 
    w/o In-Chunk Negatives & 53.9$_{\textcolor{red}{\downarrow2.7}}$ \\ 
    w/o In-Document Negatives & 53.2$_{\textcolor{red}{\downarrow3.4}}$\\ 
    w/o CMCL Loss ($\hookrightarrow$ only $L_{\text{Batch}}$) & 49.4$_{\textcolor{red}{\downarrow7.2}}$ \\ 
    w/o LI ($\hookrightarrow$ Mean Pooling+Cosine Sim.) & 38.1$_{\textcolor{red}{\downarrow18.5}}$ \\ \midrule
\end{tabular}}
\label{tab:ablation}
\end{minipage}
\hfill
\begin{minipage}[t]{0.44\textwidth}
    \vspace{0pt} 
    \centering
\includegraphics[width=\linewidth]{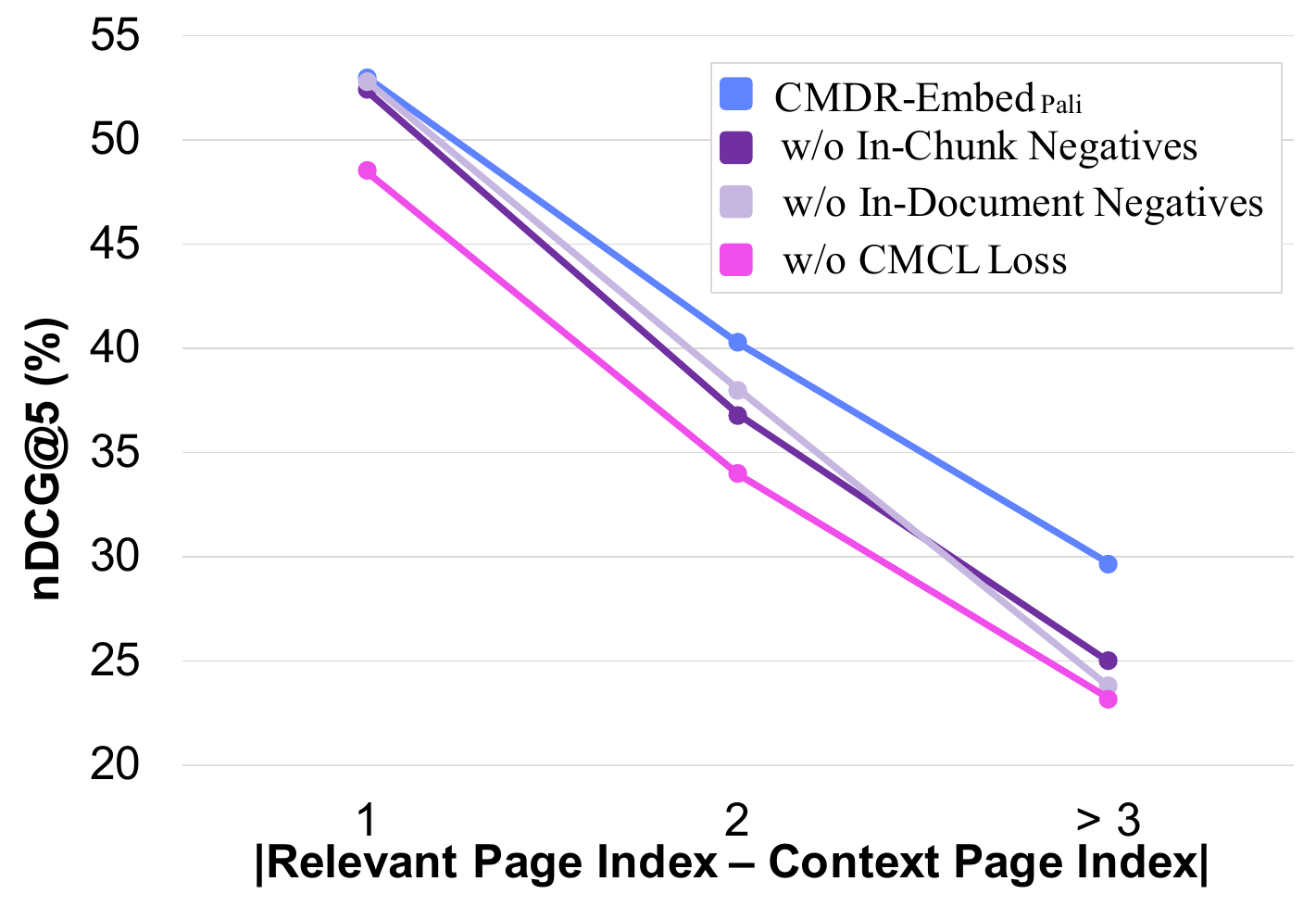}
\caption{\textbf{Comparison of our model variants using the CMCL loss} across different context lengths.}
\label{fig:rel_context_distance}
\end{minipage}
\end{figure}

\subsection{Ablation Studies}

\medskip\noindent\textbf{Effect of context-aware hard negatives and CMCL loss.} 
Table~\ref{tab:ablation} shows that the CMCL loss consistently improves performance across different backbones. In addition, both types of hard negatives contribute to these improvements.
To better understand their effects, we analyze model variants trained with the CMCL loss under different context distances, defined as the distance between the relevant page and the context page.
As shown in Figure~\ref{fig:rel_context_distance}, the two types of hard negatives exhibit complementary effects across both short- and long-context ranges. Specifically, in-chunk negatives are more effective when the relevant context appears on nearby pages, while in-document negatives mainly improve performance when the required information is located on distant pages.

\medskip\noindent\textbf{Effect of Late Interaction (LI).}
Table~\ref{tab:ablation} shows that our multi-vector embeddings with LI significantly outperform single-vector embeddings that use mean pooling over query and page representations with cosine similarity. A similar trend is observed when using last-token pooling~\cite{jiang2024vlm2vec,li2026qwen3}, which results in a 19.1-point drop in \cmdrembedpali. These results indicate that LI is particularly effective at preserving fine-grained token-level representations, enabling the model to capture both page-specific content and cross-page contextual information.

\begin{figure}[t]
\centering
\begin{minipage}[t]{0.39\textwidth}
    \vspace{0pt} 
    \centering
    \includegraphics[width=\textwidth]{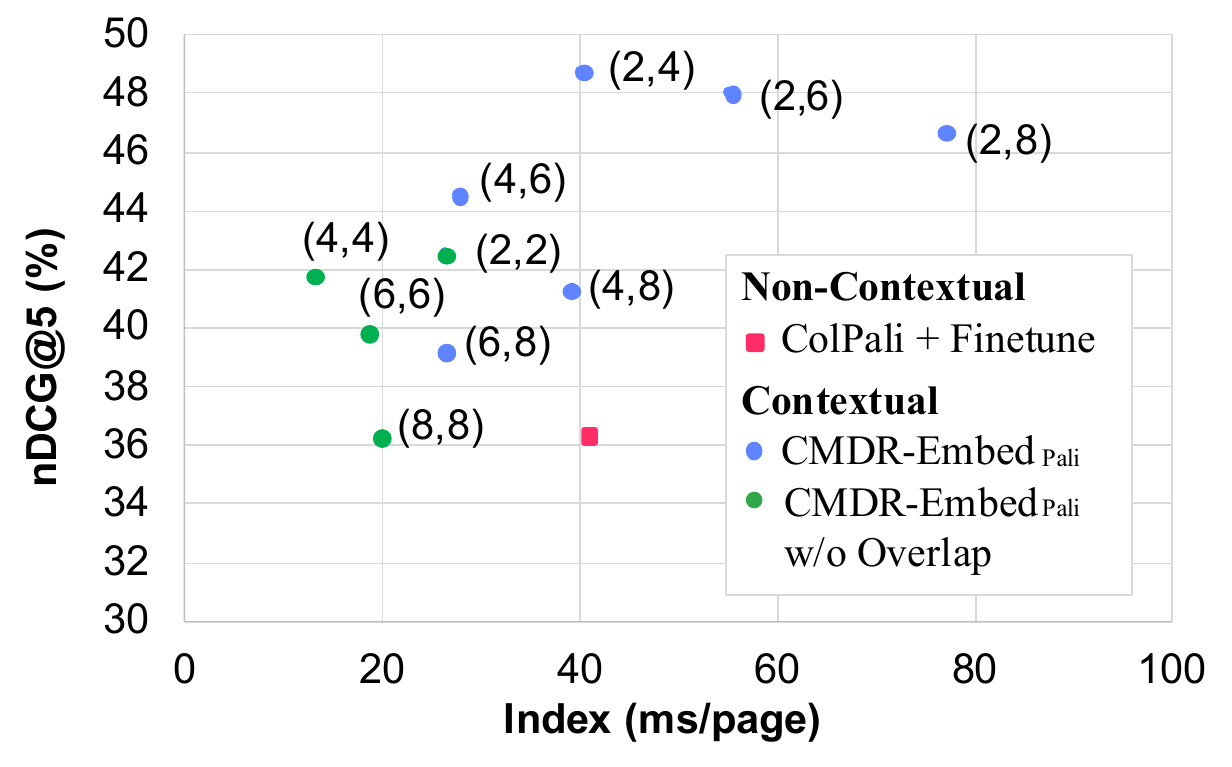}
    \captionof{figure}{\textbf{Efficiency analysis} by varying (stride $s$, window $w$).}
    \label{fig:sliding_window}
\end{minipage}
\hfill
\begin{minipage}[t]{0.575\textwidth}
    \vspace{0pt} 
    \centering
    \includegraphics[width=\textwidth]{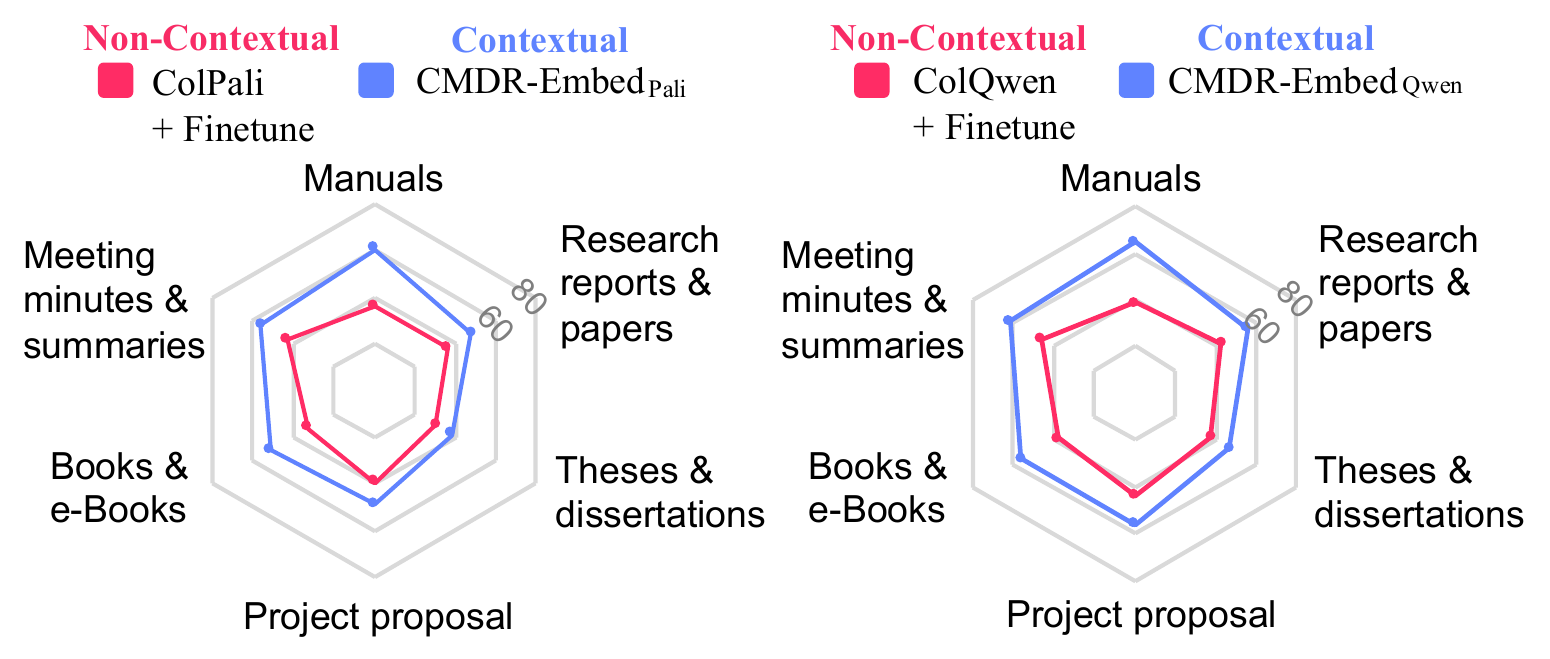}
    \captionof{figure}{\textbf{Fine-grained contextual retrieval results} under six different document types.}
    \label{fig:doctype_performance}
\end{minipage}
\end{figure}

\subsection{Further Analysis}

\medskip\noindent\textbf{How efficient is our model with a sliding window?}
Figure~\ref{fig:sliding_window} shows that our model with a stride $s$ of 2 and a window size $w$ of 4 achieves the best performance. Notably, its indexing time is comparable to that of the backbone while delivering a substantial improvement. We further observe that overlap between adjacent chunks ($s < w$) yields additional gains, indicating that sharing context across chunks is effective for capturing cross-page dependencies. However, as the overlap length $(w - s)$ increases, performance tends to decline. We hypothesize that a larger overlap produces highly similar representations for neighboring chunks, making them harder to distinguish during retrieval. Moreover, a larger overlap increases indexing costs. This highlights the importance of balancing stride and window sizes to achieve both strong performance and practical efficiency.

\medskip\noindent\textbf{On which document types do our models perform well or struggle?}
\label{rq:document_category}
As illustrated in Figure~\ref{fig:doctype_performance}, \cmdrmethodname outperforms non-contextual models across all document types. In particular, \cmdrmethodname shows significant improvements on Manuals and Books \& e-Books. This indicates that our models are especially effective at understanding documents with strong inter-page dependencies, such as procedural instructions and storytelling. However, both \cmdrmethodname and non-contextual models underperform on Research reports \& papers and Theses \& dissertations. We hypothesize that this is due to the high density of domain-specific terminology, complex figures and tables, and long-range dependencies (e.g., references), which remain challenging for current LVLM-based retrievers. 

\begin{figure}[t]
\centering
\begin{minipage}[t]{0.32\textwidth}
    \vspace{0pt} 
    \centering
\includegraphics[width=\linewidth]{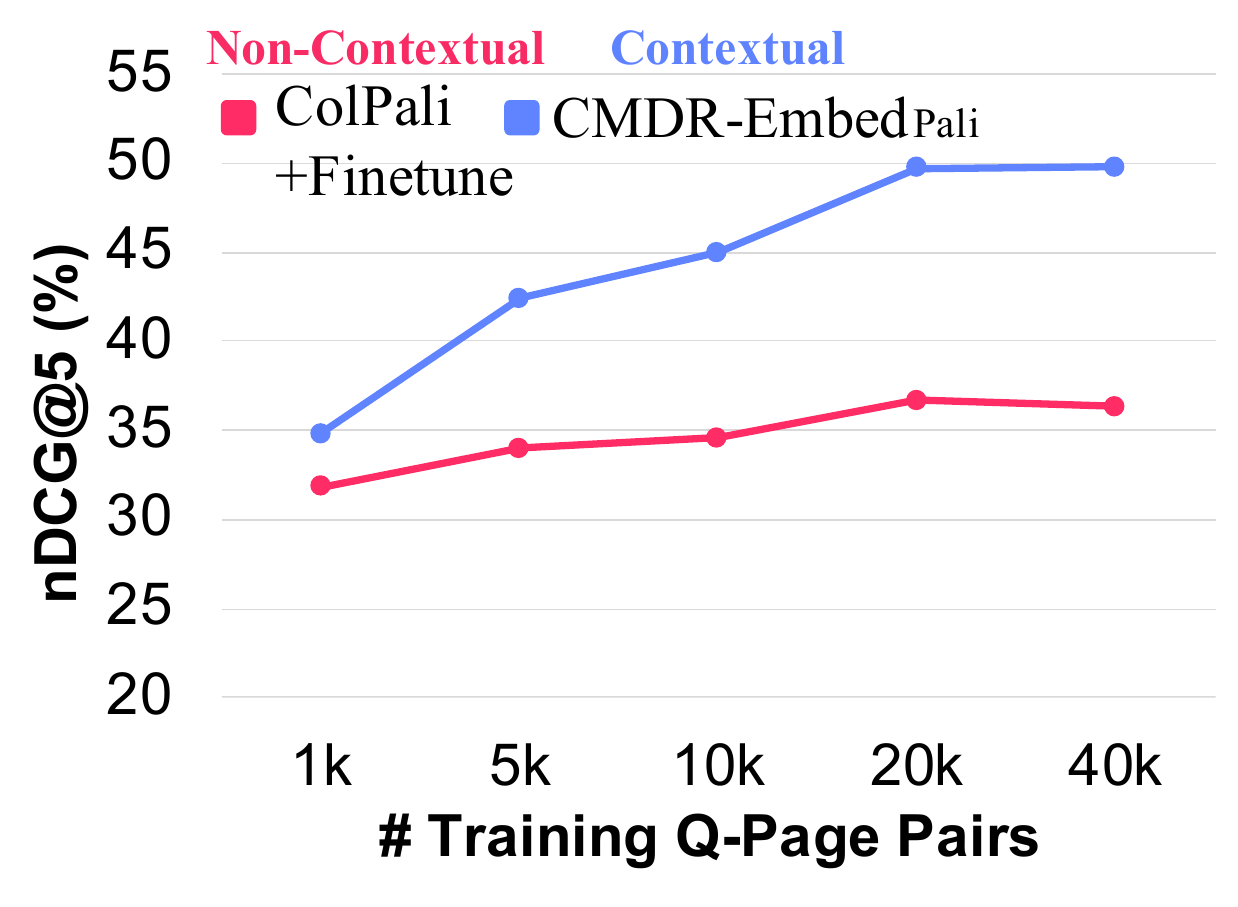}
\captionof{figure}{\textbf{Analysis of training scale} by increasing training instances.}
\label{fig:training_efficiency}
\end{minipage}
\hfill
\begin{minipage}[t]{0.32\textwidth}
    \vspace{0pt} 
    \centering
\includegraphics[width=\linewidth]{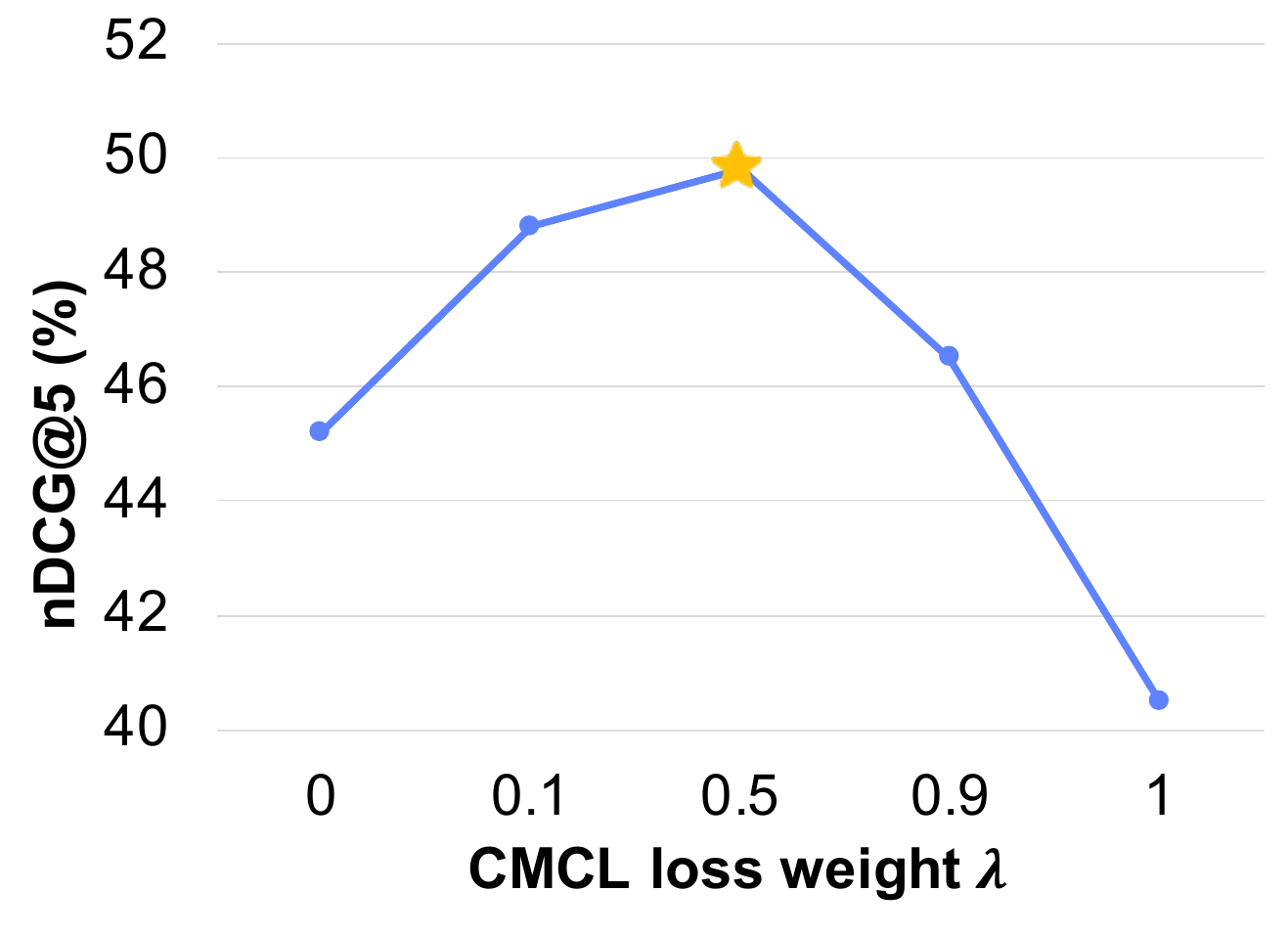}
    \captionof{figure}{\textbf{Analysis of CMCL loss weight} by varying different $\lambda$.}
\label{fig:loss_weight}
\end{minipage}
\hfill
\begin{minipage}[t]{0.32\textwidth}
    \vspace{0pt} 
    \centering
\captionof{table}{\textbf{Token pooling} for improving efficiency. A pooling factor of $1$ denotes the original performance.}
\scalebox{0.75}{
\setlength{\tabcolsep}{3pt}
\small
    \begin{tabular}{lccc}\toprule
        Pooling & Store & Search & Overall \\ 
        Factor & (GB) & (MM:SS) & (\%) \\ \midrule
        \hgrow 1  & 23.0 & 6:11 & \textbf{49.8} \\
        2 & 12.0 & 4:05 & 49.7 \\
        3 & 7.7 & 2:50 & 49.0 \\
        4 & 5.8 & 2:11 & 48.3 \\
        5 & \textbf{4.6} & \textbf{1:44} & 48.5 \\ \bottomrule
    \end{tabular}}
\label{tab:pooling}
\end{minipage}
\end{figure}

\medskip\noindent\textbf{What is the effect of training data size?}
We analyze how the scale of the training data (\cmdrsynthname) affects performance by randomly sampling subsets from the original 40k training instances and finetuning each model accordingly. Figure~\ref{fig:training_efficiency} shows that performance improves as the data size increases, with gains saturating around 20k--40k instances. Notably, the improvements are larger for \cmdrmethodname, suggesting that explicitly modeling context leverages supervision more effectively. However, the saturation trend across both model types suggests that simply scaling up the training data is unlikely to yield further improvements. This suggests that the backbone's representational capacity is the primary limiting factor, and that stronger backbones are required to improve performance.

\medskip\noindent\textbf{How does $\lambda$ in CMCL loss affect performance?} We analyze the sensitivity of $\lambda$ in \cmdrembedpali, which balances the context-aware loss $\mathcal{L}_{\text{Context}}$ and the standard in-batch contrastive loss $\mathcal{L}_{\text{Batch}}$. As shown in Figure~\ref{fig:loss_weight}, a balanced combination is crucial: If $\lambda$ is too small, the model relies mainly on in-batch negatives and does not sufficiently learn to distinguish semantically similar pages within the same document, which weakens page-level discriminability. In contrast, if $\lambda$ is too large, training is dominated by context-aware negatives, which makes the optimization problem excessively difficult and slows convergence. We find that setting the CMCL loss weight, $\lambda$, to 0.5 yields the best trade-off, balancing training convergence and overall retrieval performance.

\medskip\noindent\textbf{Can our models improve retrieval efficiency without sacrificing retrieval performance?} 
Unlike single-vector embedding models~\cite{ma2024unifying,yu2024visrag}, our model stores multiple vectors per page (one per image patch), which increases both storage requirements and retrieval latency. To address this issue, we apply hierarchical token pooling~\cite{clavie2024reducing} to \cmdrembedpali, a training-free method that aggregates redundant image patches (e.g., white background regions) and substantially reduces the number of stored embeddings while preserving most of the semantic content. As shown in Table~\ref{tab:pooling}, hierarchical token pooling reduces the total number of vectors by 80.0\% and accelerates search by 3.57$\times$, while retaining 97.3\% of the original performance with a pooling factor of 5.

\begin{table}[t!]
    \centering
    \caption{\textbf{Non-contextual and contextual retrieval results} (nDCG@5) on ViDoRe~\cite{faysse2024colpali} dataset and our \cmdrdatasetname, respectively.}
        \scalebox{0.76}{
    \tabcolsep=2pt
    \small
    \begin{tabular}{lcccccccccccc} 
        \toprule
        \multirow{2}{*}{Model} & \multicolumn{11}{c}{Non-Contextual} & Contextual \\
         & ArxQ & DocQ & InfQ & TabF & TATQ & Shift & AI & Ene. & Gov. & Health & Avg. & Overall \\   \midrule
        ColPali finetuned on ViDoRe & 79.6 & 57.7 & 81.9 & 81.5 & 65.5 & 70.0 & 97.1 & 92.1 & 92.2 & 94.1 & 81.2 & 32.4 \\
        \cmdrembedpali & 76.4 & 52.8 & 82.4 & 88.2 & 59.1 & 75.1 & 97.0 & 91.0 & 92.2 & 95.0 & 80.9 & \textbf{49.8} \\
        \hspace{0.2cm} w/ Multitask Learning & \textbf{84.2} & \textbf{60.3} & \textbf{83.5} & \textbf{90.9} & \textbf{71.6} & \textbf{78.4} & \textbf{99.4} & \textbf{95.1} & \textbf{96.2} & \textbf{96.1} & \textbf{85.6} & 47.1 \\ \midrule
        ColQwen finetuned on ViDoRe & \textbf{87.4} & 60.4 & \textbf{92.1} & 89.6 & 81.5 & \textbf{89.5} & 98.3 & \textbf{96.3} & 95.8 & 98.6 & 88.9 & 32.6 \\ 
        \cmdrembedqwen & 82.3 & 56.1 & 88.1 & 92.3 & 74.3 & 84.7 & 94.8 & 92.8 & 93.7 & 95.4 & 85.5 & \textbf{56.6} \\
        \hspace{0.2cm} w/ Multitask Learning & 87.2 & \textbf{63.7} & 89.7 & \textbf{91.0} & \textbf{81.8} & 88.3 & \textbf{98.6} & 95.5 & \textbf{97.1} & \textbf{98.9} & \textbf{89.2} & 56.1\\
        \bottomrule
    \end{tabular}
    }
    \label{tab:non_context}
\end{table}

\medskip\noindent\textbf{Can our models generalize effectively to non-contextual retrieval?}
To improve performance on non-contextual retrieval without degrading contextual retrieval capability, we finetune \cmdrmethodname with a multitask learning strategy on both contextual and non-contextual retrieval tasks, using \cmdrsynthname and ViDoRe~\cite{faysse2024colpali} datasets. This strategy enables the model to jointly learn document contextual reasoning and page-level relevance matching. 
Table~\ref{tab:non_context} shows that our multitask learning improves non-contextual retrieval by an average of 4.7\% for \cmdrembedpali, while maintaining comparable performance on contextual retrieval. These results indicate that multitask learning can effectively generalize across both contextual and non-contextual retrieval settings.

\begin{figure*}[t!]
    \centering
\includegraphics[width=.99\textwidth]{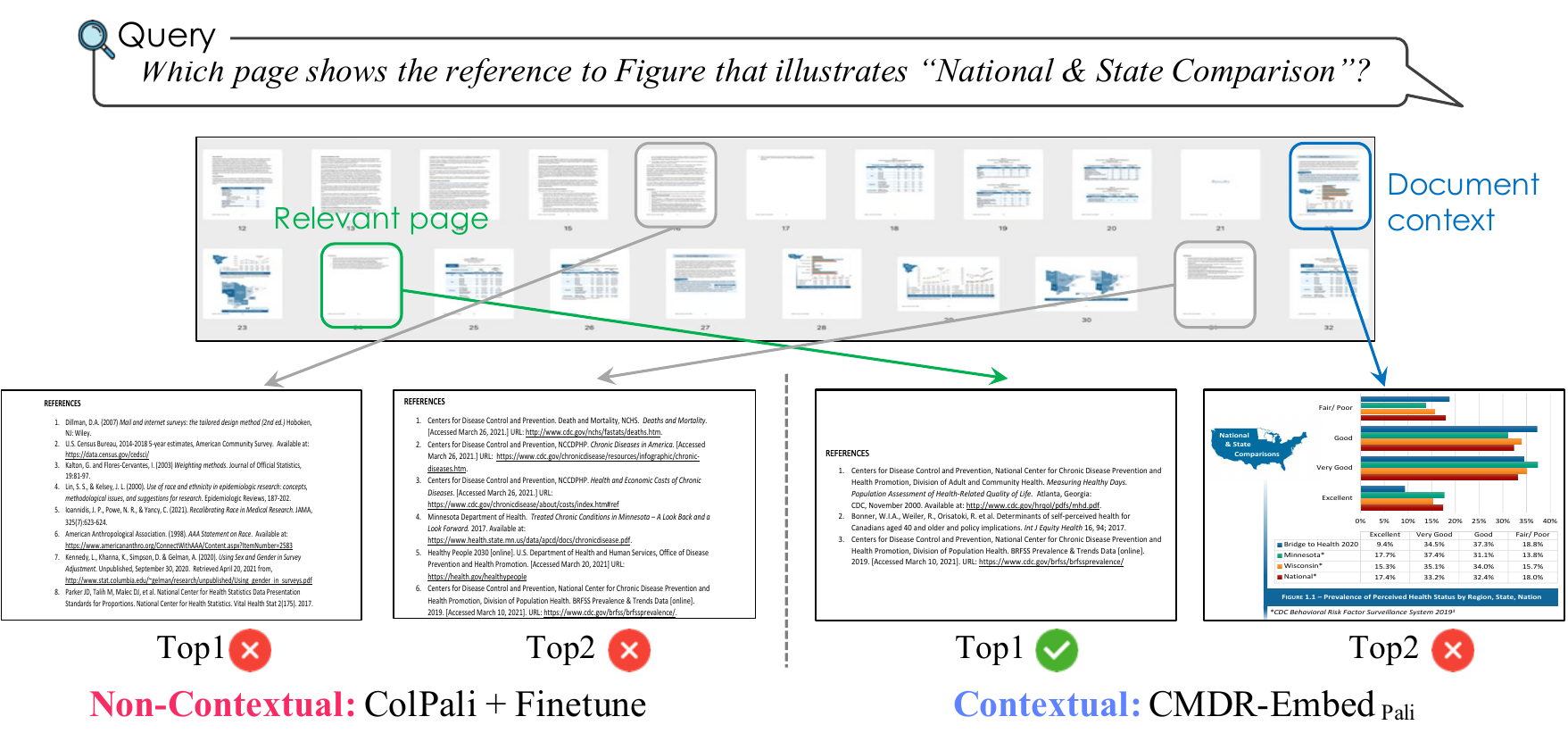}
    \caption{\textbf{Qualitative results} of \cmdrmethodname compared to the non-contextual model.}
    \label{fig:qualitative_results}
\end{figure*}

\medskip\noindent\textbf{Qualitative analysis.} We illustrate the advantages of our contextual embedding approach in Figure~\ref{fig:qualitative_results}. \cmdrmethodname successfully retrieves the correct page for a coreference resolution query that requires reasoning across multiple pages within a document. In contrast, the non-contextual baseline (ColPali) fails to capture cross-page relationships and instead retrieves incorrect pages, relying on a superficial lexical match to the term ``\textit{reference}'' in the query. These results highlight the importance of modeling document context to resolve ambiguities that page-level matching alone cannot address.

\medskip\noindent\textbf{Error analysis.}
To further investigate failure cases, we manually analyzed incorrectly retrieved examples and categorized the main errors into three types (see the Supplementary Material for representative examples) as follows:

\begin{itemize}
\item \textbf{Long-context understanding} errors occur when the model fails to capture long-range dependencies across multiple pages, such as coreference resolution and multi-hop reasoning. These errors are particularly prevalent in document types with strong inter-page dependencies (e.g., thesis \& dissertations, reports \& papers). Expanding chunk length while maintaining efficient indexing and retrieval would be promising.

\item \textbf{Subsequent-context understanding} errors occur when the model fails to utilize contextual information that appears \textit{after} the relevant page in the document. As our model relies on causal attention due to backbone constraints, it cannot directly attend to information on later pages during encoding. Adapting bidirectional attention by training on a large-scale dataset is a promising approach to overcome this limitation and enable the model to incorporate contextual information from subsequent pages.

\item \textbf{Fine-grained visual understanding} errors arise when the information is embedded in complex figures, tables, or charts that require fine-grained visual grounding. In such cases, the model must accurately align textual queries with specific visual elements, such as small labels, cell values, or chart legends. Improving the encoder’s ability to represent fine-grained visual information and incorporating stronger cross-modal alignment mechanisms would be promising directions.
\end{itemize}

\section{Conclusion}
\label{sec:conclusion}
We introduced a new multimodal document retrieval task and benchmark, \cmdrtaskname and \cmdrdatasetname, to evaluate retrieval systems that require contextual reasoning across multiple pages. Unlike existing benchmarks that focus on page-level lexical or semantic matching, our benchmark explicitly requires modeling cross-page dependencies. To address this challenge, we proposed \cmdrmethodname, a contextual multimodal embedding model trained with a new contrastive learning framework, CMCL, to balance contextual modeling and page-level discriminability. Extensive experiments demonstrate that our approach significantly outperforms non-contextual embedding models, highlighting that modeling context is essential for advancing multimodal document retrieval. Therefore, future multimodal document retrieval systems should move beyond isolated page representations and explicitly model document-level context. We hope that \cmdrdatasetname will serve as a strong testbed for future research on context-aware multimodal embeddings, RAG, and long-context document understanding, enabling more reliable document retrieval over real-world documents.

\medskip\noindent\textbf{Limitations.} 
We identify three primary limitations of the current benchmark and model. First, although \cmdrdatasetname is carefully constructed with high-quality human annotations, the number of newly annotated queries in \cmdrdatasetname is comparable to that of the most closely related benchmark, MMLB-Doc~\cite{ma2024mmlongbench} (800 queries in \cmdrdatasetname vs. 898 queries in MMLB-Doc). In the information retrieval community, 30–50 queries per category, substantially fewer than in \cmdrdatasetname, are often considered sufficient for reliable evaluation, as each query requires annotation across numerous documents and pages for relevance~\cite{weller2025followir,webber2008statistical}. 
Second, our model still inherits architectural constraints from the backbone. Most LVLMs rely on causal attention, which restricts bidirectional context modeling across pages; adopting bidirectional attention could enable richer document-level interactions. Third, the jointly processable context length is constrained by memory and computational limits, limiting the length of very long image sequences. Extending the context length while maintaining efficiency remains an important direction for future work.

\bibliographystyle{splncs04}
\bibliography{main}

\newpage

\appendix

{
\centering
\Large
\textbf{CMDR:\\Contextual Multimodal Document Retrieval}\\
\vspace{0.5em}Supplementary Material \\
}

\renewcommand\thefigure{\Alph{section}.\arabic{figure}}
\renewcommand\thetable{\Alph{section}.\arabic{table}} 

\setcounter{figure}{0}
\setcounter{table}{0}

\newcommand\beginsupplement{%
        \setcounter{table}{0}
        \renewcommand{\thetable}{\Alph{table}}%
        \setcounter{figure}{0}
        \renewcommand{\thefigure}{\Alph{figure}}%

     }

\section{\cmdrdatasetname Details}

\begin{figure}[h]
 \begin{minipage}{0.47\textwidth}
    \centering
    \captionof{table}{\textbf{Main statistics} of documents and queries in \cmdrdatasetname.}
        \scalebox{0.9}{
    \tabcolsep=1.0pt
    \small
    \begin{tabular}{lc} 
        \toprule
        Statistics & Number \\ \midrule
        Total Documents & 255 \\   
        - Manuals & 64 (25.1\%) \\ 
        - Research reports \& papers & 34 (13.3\%) \\ 
        - Theses \& dissertations & 45 (17.6\%) \\ 
        - Project proposals & 36 (14.1\%) \\ 
        - Books \& e-books & 41 (16.1\%) \\ 
        - Meeting minutes \& summaries & 35 (13.8\%) \\ 
        Total Pages (Images) & 46,781 \\ 
        Maximum Pages (Images) & 714 \\ 
        Minimum Pages (Images) & 100 \\ 
        Average Pages (Images) & 183.5 \\ \midrule
        Total Queries & 800 \\ 
        - Text Completion  & 168 (21.0\%) \\ 
        - Coreference Resolution & 181 (22.6\%) \\ 
        - Structured Understanding  & 204 (25.5\%) \\
        - Multi-hop Reasoning  & 247 (30.9\%) \\ 
        Maximum Query Length & 52 \\
        Minimum Query Length & 5 \\
        Average Query Length & 20.0 \\
        \bottomrule
    \end{tabular}
    }
    \label{tab:stat}
\end{minipage}
\begin{minipage}{0.49\textwidth}
        \begin{minipage}{\linewidth}
            \centering
            \begin{subfigure}{0.77\linewidth}
                \includegraphics[width=\linewidth, trim={0.7em 0 0.6em 0em}, clip]{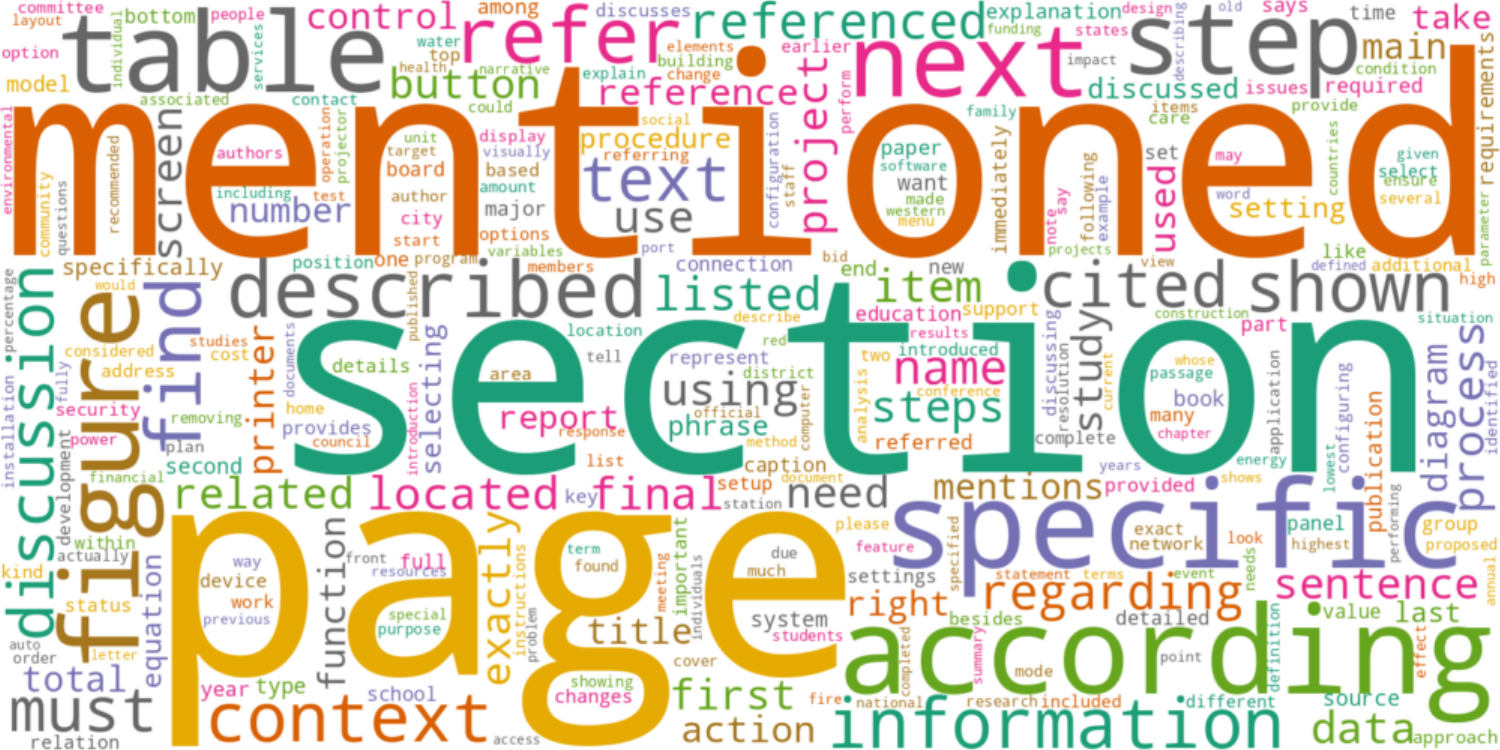}
                \caption{Word cloud of queries.}
                \label{fig:wordcloud}
            \end{subfigure}
            \begin{subfigure}{0.77\linewidth}
                \includegraphics[width=\linewidth, trim={0.7em 0 0.6em 0em}, clip]{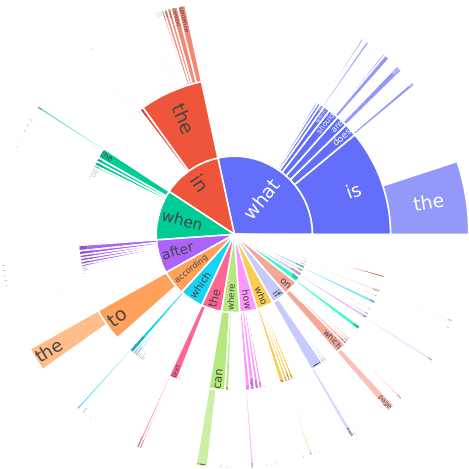}
                \caption{First three words of queries.}
                \label{fig:sunburst}
            \end{subfigure}
            \captionof{figure}{\textbf{Query distribution}.}
            \label{fig:query_distribution}
        \end{minipage}
    \end{minipage}
\end{figure}

\medskip\noindent\textbf{Dataset statistics.} The main statistics of \cmdrdatasetname are presented in Table~\ref{tab:stat}. It contains 800 queries across 255 documents, spanning four query types and six document types. Figure~\ref{fig:query_distribution}\textcolor{red}{a} presents word clouds of the most frequently appearing words in the query texts, illustrating that \cmdrdatasetname covers a wide range of topics and words. This observation is further supported by Figure~\ref{fig:query_distribution}\textcolor{red}{b}, which is a sunburst of the first three words of the questions.

\begin{figure*}
    \centering
\includegraphics[width=.85\textwidth]{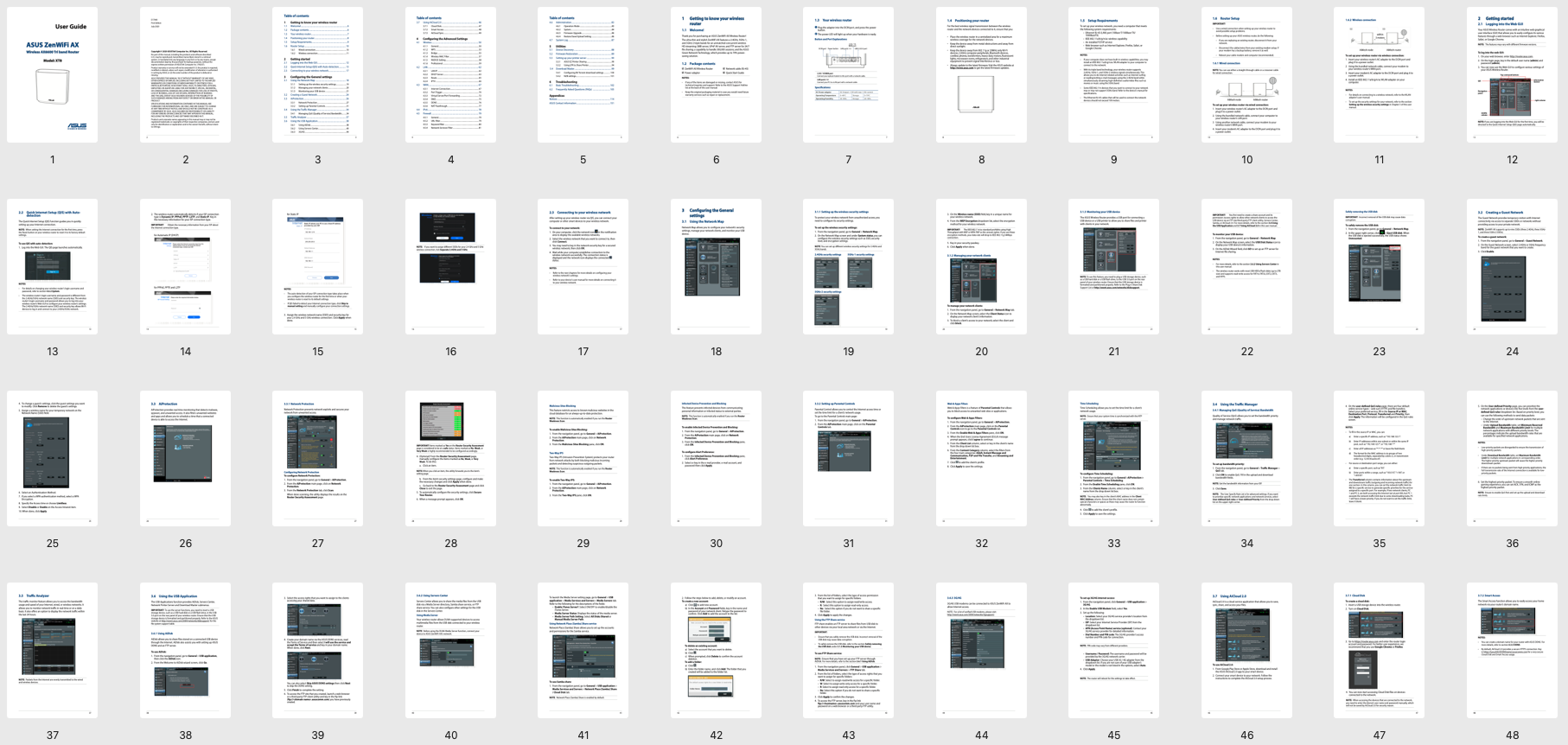}
    \caption{Document example of \textbf{Manuals}.}
    \label{fig:example_manual}
\end{figure*}

\begin{figure*}
    \centering
\includegraphics[width=.85\textwidth]{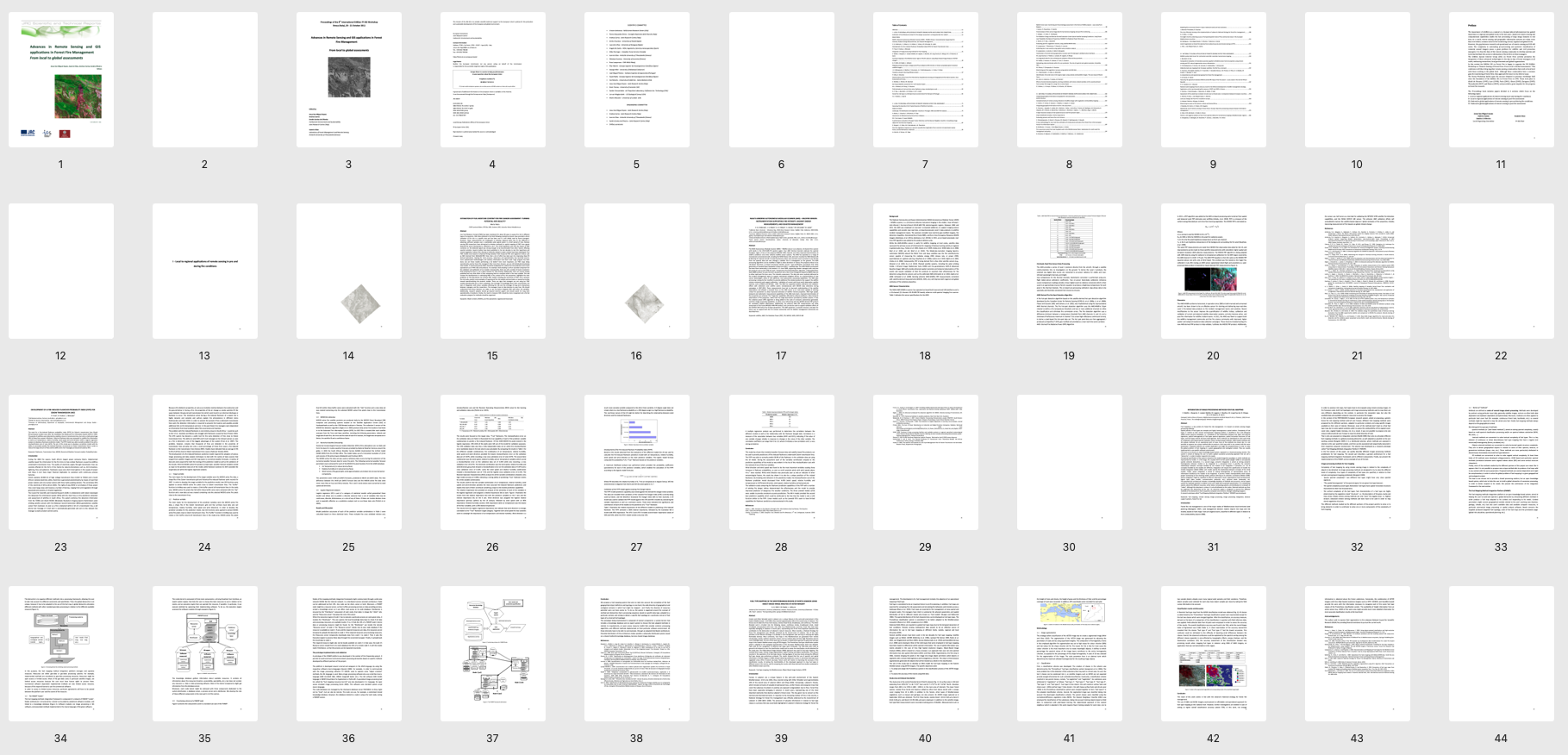}
    \caption{Document example of \textbf{Research reports \& papers}.}
    \label{fig:example_research_reports}
\end{figure*}

\begin{figure*}
    \centering
\includegraphics[width=.85\textwidth]{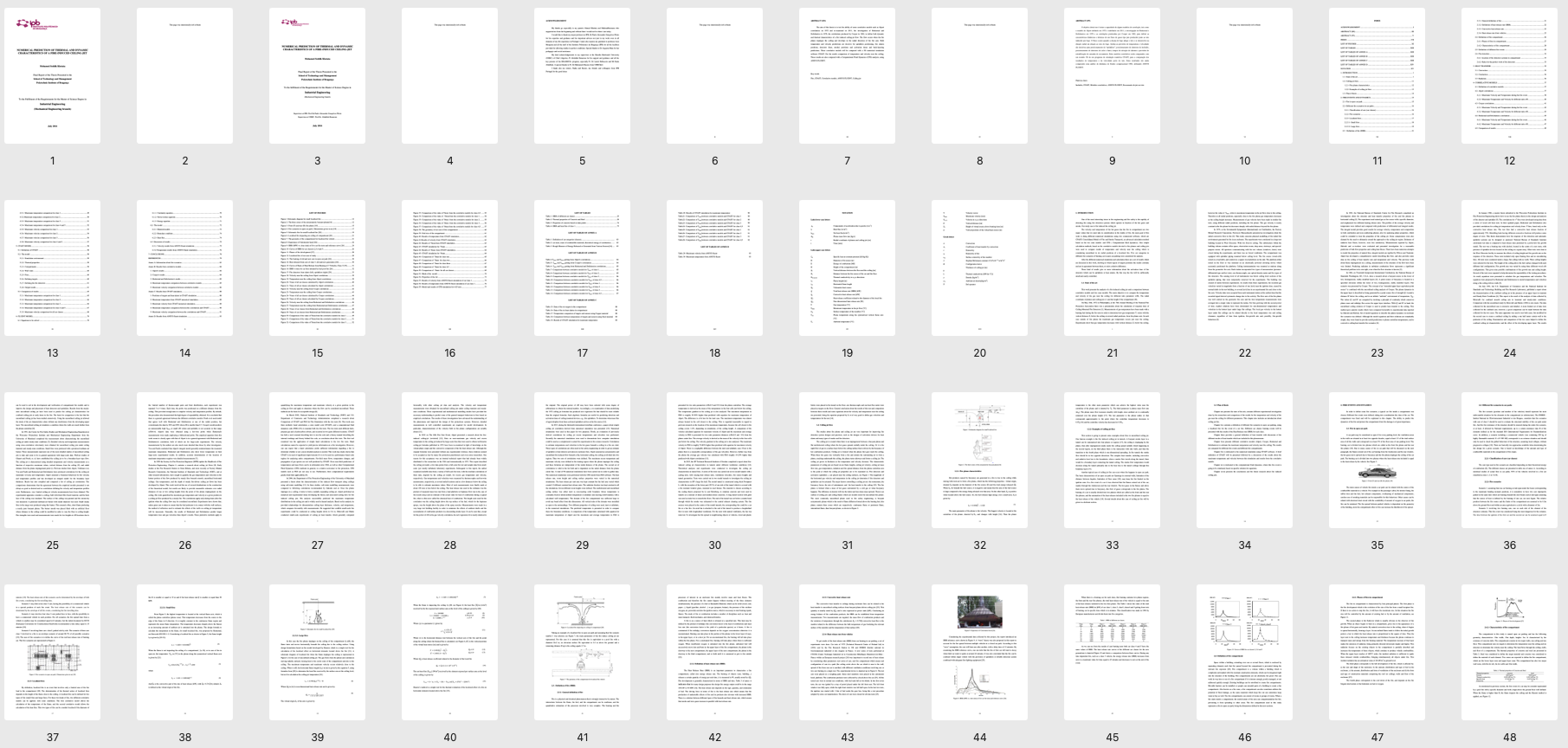}
    \caption{Document example of \textbf{Theses \& dissertations}.}
    \label{fig:example_thesis}
\end{figure*}

\begin{figure*}
    \centering
\includegraphics[width=.85\textwidth]{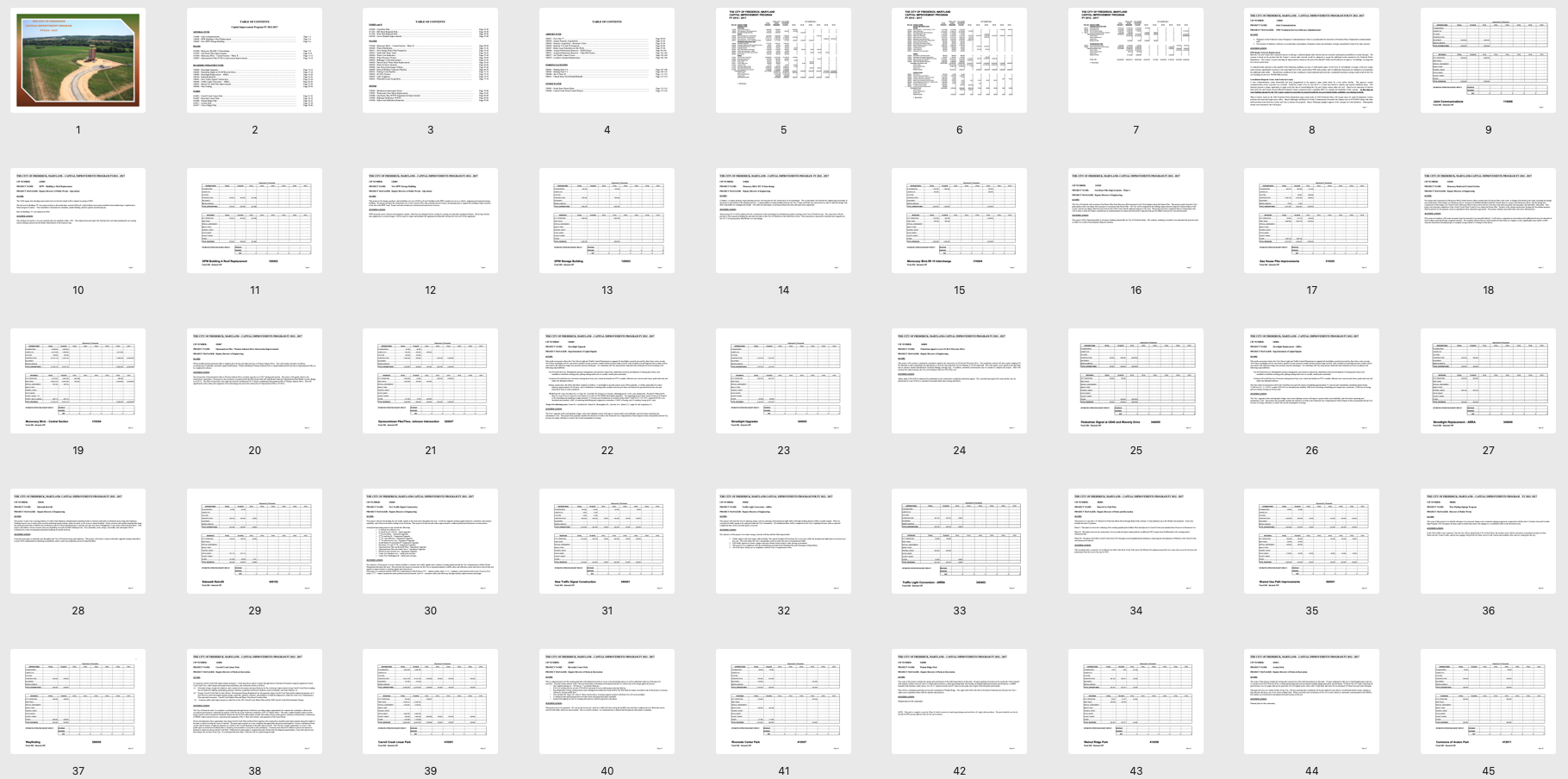}
    \caption{Document example of \textbf{Project proposals}.}
    \label{fig:example_project_proposals}
\end{figure*}

\begin{figure*}
    \centering
\includegraphics[width=.85\textwidth]{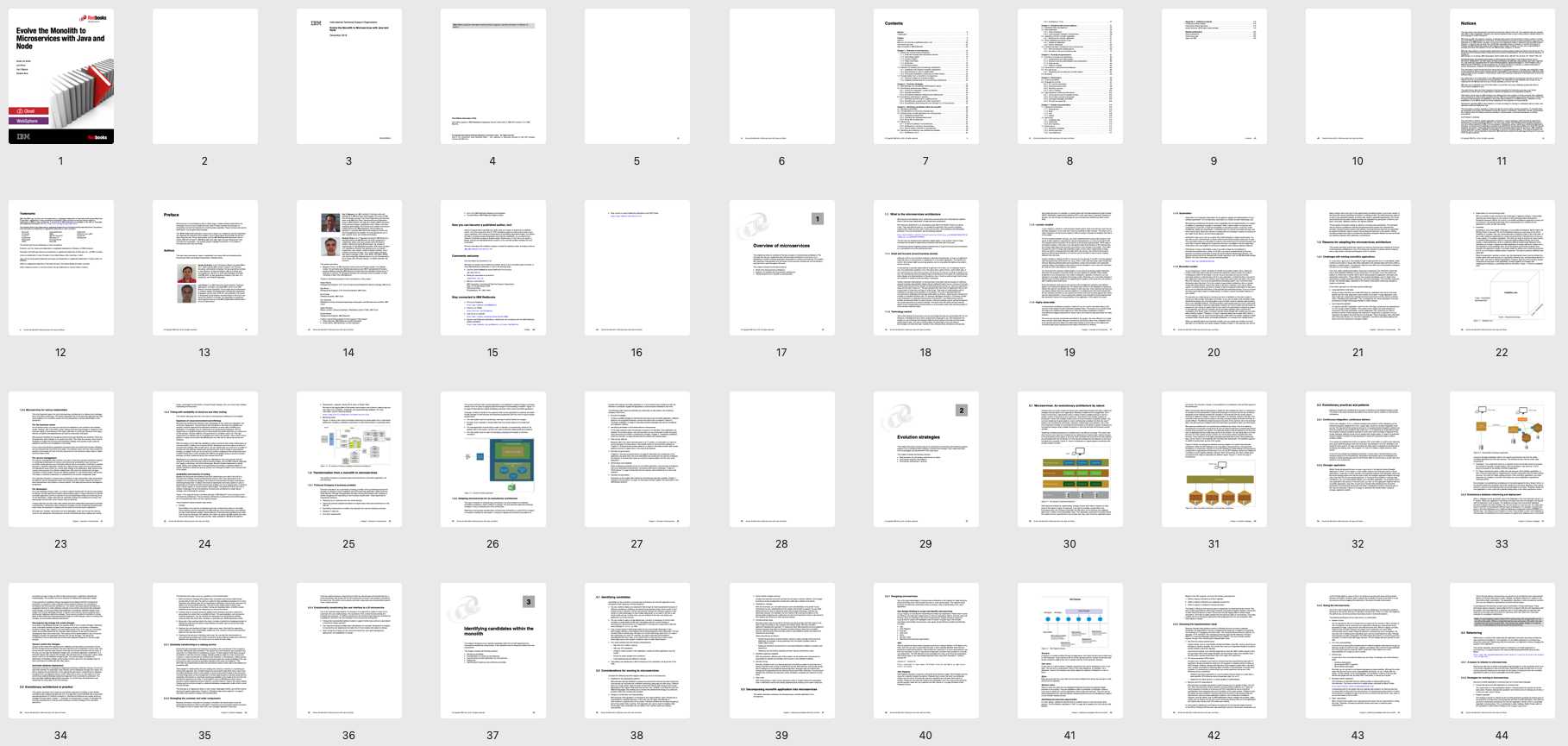}
    \caption{Document example of \textbf{Books \& e-books}.}
    \label{fig:example_books}
\end{figure*}

\begin{figure*}
    \centering
\includegraphics[width=.85\textwidth]{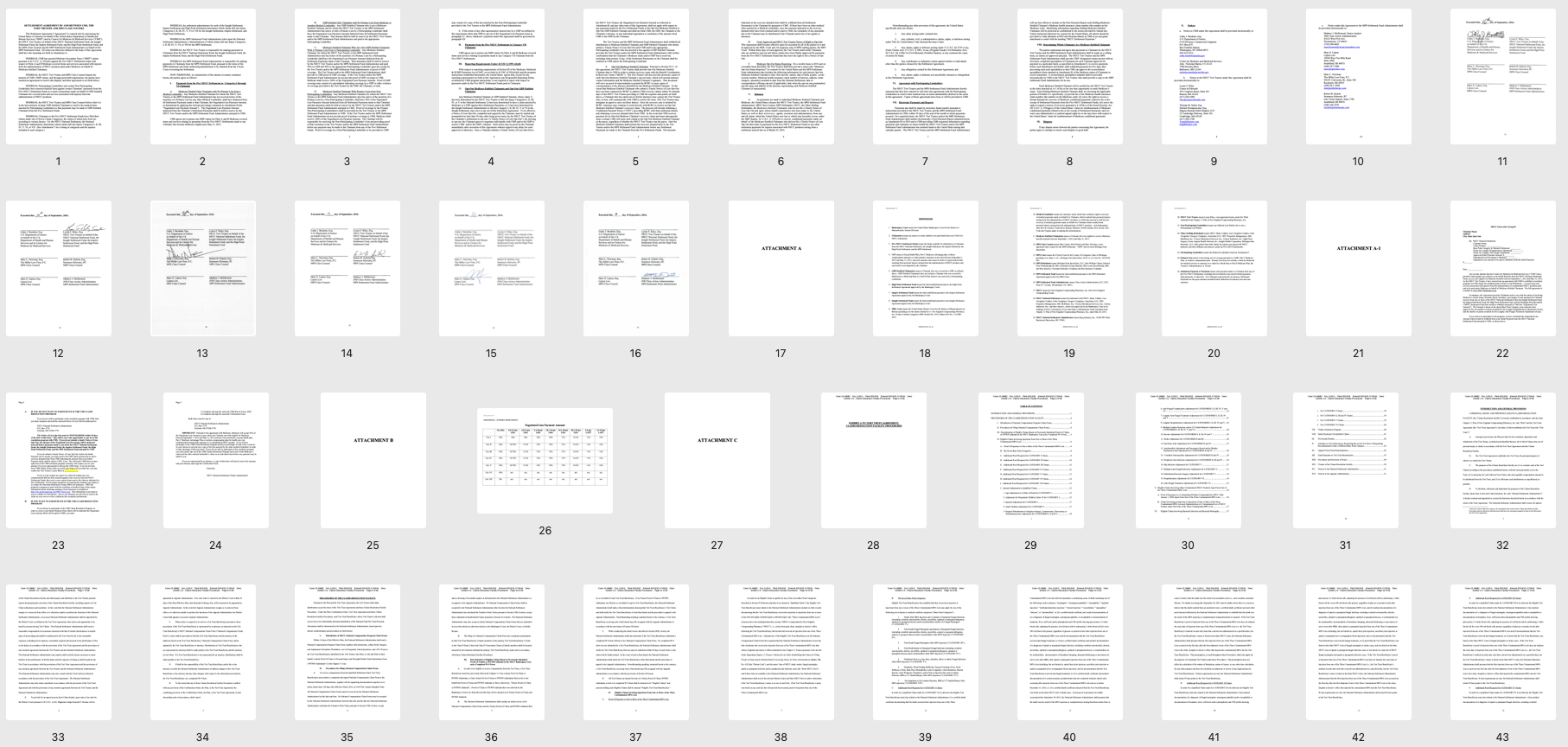}
    \caption{Document example of \textbf{Meeting minutes \& summaries}.}
    \label{fig:example_meeting_minutes}
\end{figure*}

\medskip\noindent\textbf{Document examples.}
As stated in Section~\ref{para:document_lategory}, the documents in \cmdrdatasetname can be categorized into six
types. Figure~\ref{fig:example_manual}-\ref{fig:example_meeting_minutes} show the examples of collected documents.

\begin{figure*}
    \centering
\includegraphics[width=1.\textwidth]{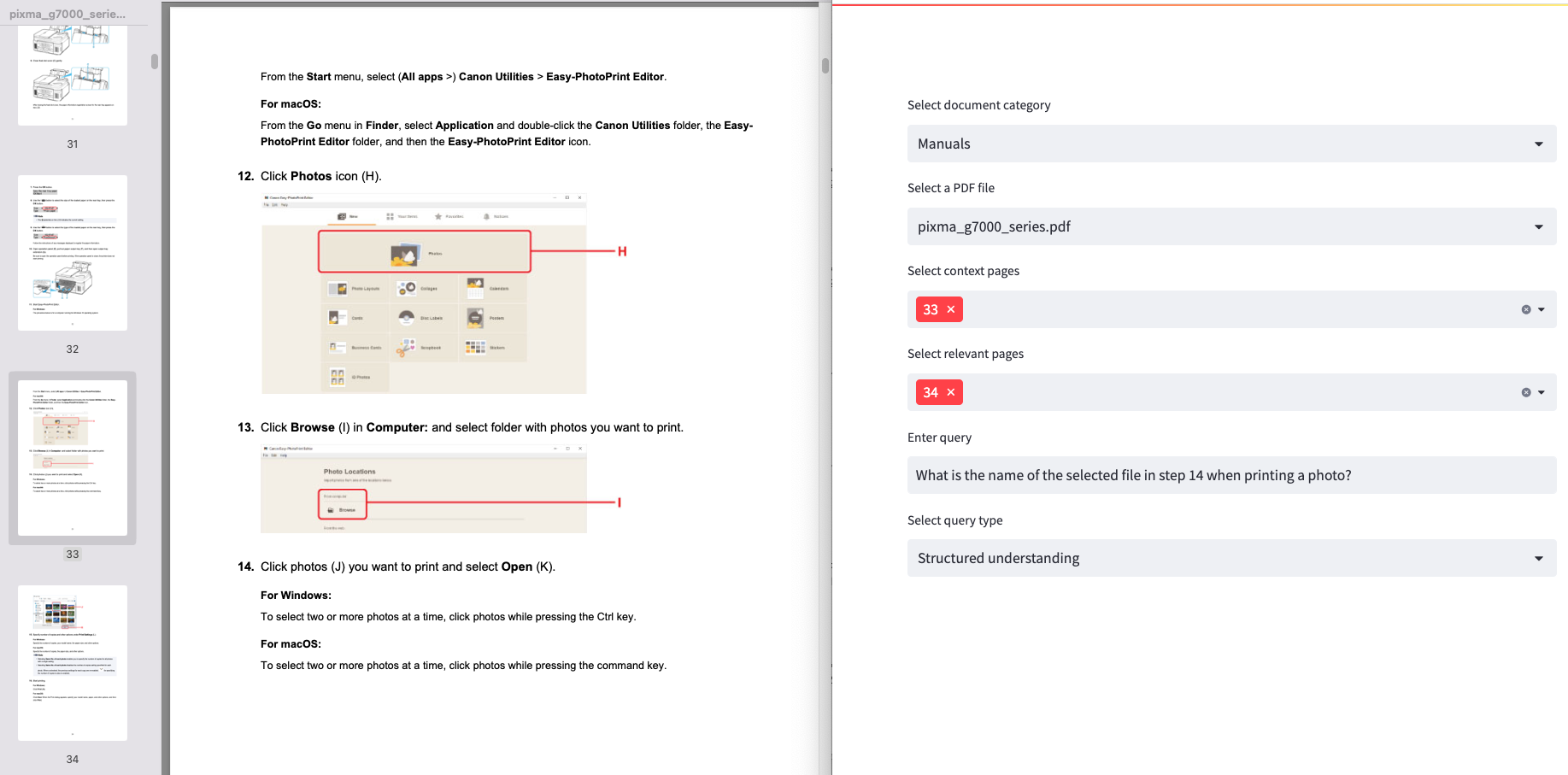}
    \caption{\textbf{GUI screenshot for annotating queries}. Before annotating queries, annotators carefully read each page to understand its structure and dependencies.}
    \label{fig:annoation_gui}
\end{figure*}

\medskip\noindent\textbf{Annotation GUI.}
Figure~\ref{fig:annoation_gui} shows the screenshot for annotating queries. The annotation GUI was implemented with Streamlit\footnote{https://streamlit.io/}.

\section{\cmdrsynthname Construction Details}
As described in Section~\ref{para:training_dataset}, we created \cmdrsynthname for training contextual multimodal embedding models through four steps as follows: 

\medskip\noindent\textbf{(1) Filtering self-contained pages.} 
Given the prompt below and a single-page image from a multi-page PDF, we feed them into Qwen2.5-VL 72B~\cite{bai2025qwen2}, which outputs a binary label indicating whether the page is self-contained.
\begin{table}
\centering
\begin{tcolorbox}[fontupper=\scriptsize, title=\footnotesize Prompt: (1) Filtering self-contained pages]
\begin{verbatim}
You are given an image of a single page from a multi-page PDF document. 
Determine whether the content of this page is **self-contained** 
(can be fully understood on its own) or **not self-contained** (requires other pages).  

Criteria for judgment:  
- Self-contained (Yes):  
  - The page presents complete information that can be understood 
    without referring to other pages.  
  - The text begins and ends naturally (not cut off in the middle).  
  - Title pages or pages with no text are considered self-contained.
  - Figures, tables, or explanations are understandable on this page alone.    
- Not self-contained (No):  
  - The page refers to figures, tables, or sections on other pages.  
  - The page is incomplete or depends on missing context.  
  - The page starts in the middle of a sentence, paragraph, or table.  
  
Output Format:  
Self-contained: [Yes/No] 
\end{verbatim}
\end{tcolorbox}
\end{table}

\medskip\noindent\textbf{(2) Retrieving context pages.}
Given the prompt below and the image classified as ``No'' (i.e., ``Not self-contained''), we feed them into UniSE~\cite{liu2025any}, a general multimodal embedding model that can take both images and text as input. Then, UniSE identifies and retrieves the page that provides essential contextual information for better understanding the given page.

\begin{table}
\centering
\begin{tcolorbox}[fontupper=\scriptsize, title=\footnotesize Prompt: (2) Retrieving context pages]
\begin{verbatim}
Retrieve the page that provides important contextual information for 
better understanding the given page.   
\end{verbatim}
\end{tcolorbox}
\end{table}

\medskip\noindent\textbf{(3) Generating contextual queries.}
Given the prompt below, $\mathtt{Image\ 1}$, which is the context page image obtained in step 2, and $\mathtt{Image\ 2}$, which is the relevant page image obtained in step 1, we feed them into Qwen2.5-VL 72B to generate a contextual retrieval-oriented query in English. The prompt in $\mathtt{\{query\ type\}}$ refers to the query types and their descriptions defined in Section~\ref{para:query_lategory}.

\begin{table}
\centering
\begin{tcolorbox}[fontupper=\scriptsize, title=\footnotesize Prompt: (3) Generating contextual queries]
\begin{verbatim}
Create one short retrieval-oriented query in English following the conditions below.

1. The query should appear to be about Image 1, but the correct answer is 
 found only in Image 2.  
  - At first glance, it should seem that the answer could be in Image 1.  
  - Only by reading both pages carefully should it become clear that the answer 
  is actually in Image 2.
  
2. The query must require contextual understanding across both pages, 
   not a simple keyword search.
   
3. Query Type: {query type}

4. Do not quote or copy any text from Image 2 directly into the query.  
  - Do not include explicit hints or keywords that would guide the reader 
  to look into Image 2.
  
5. Only output the query in the specified format without any additional explanation.

6. Use only one question word and keep the query in a single sentence.

Output Format:    
Query: <Your query here> 
\end{verbatim}
\end{tcolorbox}
\end{table}

\medskip\noindent\textbf{(4) Filtering low-quality queries.} 
Given the prompt below, the context page image ($\mathtt{Image\ 1}$), and the relevant page image ($\mathtt{Image\ 2}$), we feed them into Qwen2.5-VL 72B to identify and filter out low-quality or rule-violating queries.
\begin{table}[ht!]
\centering
\begin{tcolorbox}[fontupper=\scriptsize, title=\footnotesize Prompt: (4) Filtering low-quality queries]
\begin{verbatim}
You are an expert evaluator for multimodal document retrieval tasks.


You are given:
- Image 1
- Image 2
- Query (generated)


Your task is to determine whether the Query satisfies ALL of the following conditions.


Evaluation Criteria:

1. Apparent Focus on Image 1
   - The query should initially appear to ask about Image 1.
   - It should be plausible that a reader might expect the answer to be
   found in Image 1 at first glance.

2. Answer Actually Located in Image 2
   - The correct answer must be found only in Image 2.
   - Image 1 alone must be insufficient to answer the query correctly.

3. Cross-Page Contextual Reasoning Required
   - The query must require integrating contextual information 
   across both images.
   - The answer should NOT be obtainable via simple keyword matching 
   or surface-level lookup.

4. No Leakage from Image 2
   - The query must NOT quote or directly copy text from Image 2.
   - The query must NOT contain explicit hints or keywords that clearly 
   direct the reader to Image 2.

5. Single Question Word & Single Sentence
  - The query must:
    - Contain exactly one question word (What, Why, When, Where, Who, How).
    - Be written as a single sentence.
     
6. Query Type Consistency
   - The query must match the specified Query Type: {query_type}.
   - If it does not match the intended reasoning type, mark it as invalid.


Output Format:

Valid: Yes or No

Reasoning:
- Criterion 1: Pass or Fail – brief explanation
- Criterion 2: Pass or Fail – brief explanation
- Criterion 3: Pass or Fail – brief explanation
- Criterion 4: Pass or Fail – brief explanation
- Criterion 5: Pass or Fail – brief explanation
- Criterion 6: Pass or Fail – brief explanation

Final Verdict: Valid or Invalid


Be strict. If any single criterion fails, the Final Verdict must be Invalid. 
\end{verbatim}
\end{tcolorbox}
\end{table}

\section{Experimental Details}

\medskip\noindent\textbf{Model URLs.}
Table~\ref{tab:model_urls} shows model URLs stored in GitHub and HuggingFace used in our experiments.

\begin{table}
    \centering
    \caption{\textbf{Model URLs} on GitHub and HuggingFace.}
    \label{tab:model_urls}
        \scalebox{0.84}{
    \tabcolsep=3pt
    \small
    \begin{tabular}{ll} 
        \toprule
        Model & Model URL \\ \midrule
        BM25~\cite{lu2024bm25s} & \url{https://github.com/xhluca/bm25s} \\
        Contriever~\cite{izacard2021unsupervised} & \url{https://huggingface.co/facebook/contriever-msmarco}  \\ 
        BGE~\cite{xiao2024c} & \url{https://huggingface.co/BAAI/bge-large-en-v1.5} \\
        E5~\cite{wang2022text} & \url{https://huggingface.co/intfloat/e5-large-v2} \\
        NV-Embed-v2~\cite{lee2025nv} & \url{https://huggingface.co/nvidia/NV-Embed-v2} \\
        CLIP~\cite{radford2021learning} & \url{https://huggingface.co/openai/clip-vit-large-patch14} \\
        SigLIP~\cite{zhai2023sigmoid} & \url{https://huggingface.co/google/siglip-so400m-patch14-384} \\
        E5-V~\cite{jiang2024e5} & \url{https://huggingface.co/intfloat/e5-large} \\
        VLM2Vec~\cite{jiang2024vlm2vec} & \url{https://huggingface.co/TIGER-Lab/VLM2Vec-Full} \\
        GME~\cite{Zhang_2025_CVPR} & \url{https://huggingface.co/Alibaba-NLP/gme-Qwen2-VL-2B-Instruct} \\
        Qwen3-VL Embedding~\cite{li2026qwen3} & \url{https://huggingface.co/Qwen/Qwen3-VL-Embedding-8B} \\
        Qwen3-VL Reranker~\cite{li2026qwen3} & \url{https://huggingface.co/Qwen/Qwen3-VL-Reranker-2B} \\
        ColMobernVBERT~\cite{teiletche2025modernvbert} & \url{https://huggingface.co/ModernVBERT/colmodernvbert} \\
        DSE~\cite{ma2024unifying} & \url{https://huggingface.co/Tevatron/dse-phi3-v1.0} \\
        VisRAG-Ret~\cite{yu2024visrag} & \url{https://huggingface.co/openbmb/VisRAG-Ret} \\
        UniSE~\cite{liu2025any} & \url{https://huggingface.co/marsh123/UniSE-MLLM} \\
        ColPali~\cite{faysse2024colpali} & \url{https://huggingface.co/vidore/colpali-v1.1} \\
        ColQwen~\cite{faysse2024colpali} & \url{https://huggingface.co/vidore/colqwen2-v1.0} \\
        \bottomrule
    \end{tabular}
    }
\end{table}

\begin{table*}[t!]
    \centering
    \caption{\textbf{Contextual retrieval results} (Recall@5) on \cmdrdatasetname. The performance gain in \textcolor{darkgreen}{green} is compared to the same finetuned backbone. Overall performance reports the average scores across four query categories defined in Section~\ref{para:query_lategory}.} 
    \label{tab:retrieval_recall}
        \scalebox{0.78}{
    \tabcolsep=2.1pt
    \small
    \begin{tabular}{llllllll} 
        \toprule
        \multirow{2}{*}{Model} & \multirow{2}{*}{Backbone} & \multirow{2}{*}{\#Params} & \multicolumn{4}{c}{Query Categories} & \multirow{2}{*}{Overall} \\ 
        & & & TC & CR & SU & MR  \\
        \midrule
        \multicolumn{2}{>{\columncolor{grpB}}c}{\textbf{Non-Contextual Models}} & \multicolumn{2}{c}{\textit{Text Retrievers}} \\ 
        BM25~\cite{lu2024bm25s} & -- & -- & 42.9  & 29.3 & 32.8 & 39.7  & 36.2 \\
        Contriever~\cite{izacard2021unsupervised} & BERT-base &  109M & 53.6 & 23.8 & 36.8 & 40.5  & 38.6 \\
        BGE~\cite{xiao2024c} & BERT-base & 109M & 56.0 & 28.2 & 40.7 & 44.1 & 42.2\\
        E5~\cite{wang2022text} & BERT-large & 340M & 52.4 & 31.5 & 37.7 & 40.9  & 40.6\\
        NV-Embed-v2~\cite{lee2025nv} & Mistral-7B &  7.9B & 56.0 & 26.0  & 44.6 & 47.8 & 43.6 \\ \hdashline
        \multicolumn{8}{c}{\textit{General Multimodal Retrievers}} \\ 
        CLIP~\cite{radford2021learning} & CLIP-large & 428M & 20.2  & 7.7 & 17.6  & 20.6 & 16.6  \\
        SigLIP~\cite{zhai2023sigmoid} & SOViT-400m & 878M & 33.9  & 13.3 & 25.0 & 32.0 & 26.0  \\
        E5-V~\cite{jiang2024e5} & LLaVA-1.6 & 8.4B & 48.8  & 34.8 & 47.1  & 51.8 & 45.6  \\
        VLM2Vec~\cite{jiang2024vlm2vec} & Phi-3.5-V & 4.2B & 39.9  & 24.3 & 39.7 & 38.9 & 35.7 \\ 
        GME~\cite{Zhang_2025_CVPR} & Qwen2-VL & 2.2B & 57.7 & 38.7 & 49.0& 53.8 & 49.8 \\
        Qwen3-VL Embedding~\cite{li2026qwen3} & Qwen3-VL & 8B & 57.7 & 42.5 & 55.9 & 55.5 & 52.9 \\
        \hdashline
        \multicolumn{8}{c}{\textit{Multimodal Document Retrievers}} \\ 
        ColMobernVBERT~\cite{teiletche2025modernvbert} & ModernBERT & 250M & 48.8 & 34.8 & 39.2 & 40.9 &  40.9  \\
        DSE~\cite{ma2024unifying} & Phi-3-V & 4.2B &  54.2  & 35.9  & 47.1 & 46.6 & 45.9\\
        VisRAG-Ret~\cite{yu2024visrag} & MiniCPM-V & 3.4B & 53.6  & 30.4 & 45.6 & 50.2 & 44.9 \\
        UniSE~\cite{liu2025any} & Qwen2-VL & 2.2B & 49.4 & 37.0  & 49.0 & 49.0 & 46.1 \\
        ColPali~\cite{faysse2024colpali} & Paligemma & 2.9B & 60.1 & 37.6& 48.0& 54.3  & 50.0 \\
        \hspace{0.2cm}+Finetuned on \cmdrsynthname & Paligemma & 2.9B & 62.5 & 42.5 & 55.9& 56.3 & 54.3 \\
        ColQwen~\cite{faysse2024colpali} & Qwen2-VL & 2.2B & 61.3 & 43.1 & 52.0 & 55.1 & 52.9 \\ 
        \hspace{0.2cm}+Finetuned on \cmdrsynthname & Qwen2-VL & 2.2B & 66.1 &  50.8 & 57.8& 60.3 & 58.8\\
       \multicolumn{2}{>{\columncolor{grpA}}c}{\textbf{Contextual Models  (Ours)}} \\ 
        \cmdrembedpali & ColPali & 2.9B & 77.4$_{\textcolor{darkgreen}{\uparrow14.9}}$ & 50.3$_{\textcolor{darkgreen}{\uparrow7.8}}$ &78.9$_{\textcolor{darkgreen}{\uparrow23.0}}$ & 72.1$_{\textcolor{darkgreen}{\uparrow15.8}}$ & 69.7$_{\textcolor{darkgreen}{\uparrow15.4}}$ \\
        \cmdrembedqwen & ColQwen & 2.2B & \textbf{85.1}$_{\textcolor{darkgreen}{\uparrow19.0}}$ & \textbf{54.7}$_{\textcolor{darkgreen}{\uparrow3.9}}$ & \textbf{81.9}$_{\textcolor{darkgreen}{\uparrow24.1}}$ & \textbf{78.5}$_{\textcolor{darkgreen}{\uparrow18.2}}$ & \textbf{75.1}$_{\textcolor{darkgreen}{\uparrow16.3}}$ \\
        \bottomrule
    \end{tabular}
    }
\end{table*}

\begin{table*}[t!]
    \centering
    \caption{\textbf{Contextual retrieval with reranker} using Qwen3-VL Reranker. nDCG@5/Recall@5.} 
    \label{tab:rerank}
        \scalebox{0.88}{
    \tabcolsep=3.5pt
    \small
    \begin{tabular}{lccccc} 
        \toprule
        \multirow{2}{*}{Model}  & \multicolumn{4}{c}{Query Categories} & \multirow{2}{*}{Overall} \\ 
        & TC & CR & SU & MR  \\
        \midrule 
        ColPali~\cite{faysse2024colpali}  & 39.1/60.1 & 24.2/37.6& 30.2/48.0& 35.4/54.3  & 32.2/50.0 \\
        \hspace{0.2cm} + Finetuned on \cmdrsynthname & 41.0/62.5 & 27.5/42.5 & 36.8/55.9& 39.9/56.3 & 36.3/54.3 \\
        \hspace{0.4cm} + Qwen3-VL Reranker~\cite{li2026qwen3} & 42.5/66.7 & 35.5/50.8 & 37.6/59.8 & 42.3/61.1 & 39.5/59.6 \\ \hdashline
        \cmdrembedpali & \textbf{53.8}/\textbf{77.4} & \textbf{33.3}/\textbf{50.3} & \textbf{57.7}/\textbf{78.9} & \textbf{54.3}/\textbf{72.1} & \textbf{49.8}/\textbf{69.7} \\
        \hspace{0.4cm} + Qwen3-VL Reranker~\cite{li2026qwen3} & 44.8/71.4 & 36.9/54.7 & 38.4/62.7 & 43.7/64.8 & 41.0/63.4 \\
    \bottomrule
    \end{tabular}
    }
\end{table*}

\begin{table*}[t!]
    \centering
    \caption{\textbf{Text retrievers} using different OCR systems. nDCG@5/Recall@5.} 
    \label{tab:ocr}
        \scalebox{0.88}{
    \tabcolsep=3.5pt
    \small
    \begin{tabular}{llccccc} 
        \toprule
        \multirow{2}{*}{Embedding Model} & \multirow{2}{*}{OCR Model} & \multicolumn{4}{c}{Query Categories} & \multirow{2}{*}{Overall} \\ 
        & & TC & CR & SU & MR  \\
        \midrule 
    BGE~\cite{xiao2024c} & Tesseract~\cite{smith2007overview} & 42.8/56.0 & 18.4/28.2 & 28.5/40.7 & 29.5/44.1 & 29.8/42.2 \\               
    BGE~\cite{xiao2024c} & GOT-OCR2.0~\cite{wei2024general} & 43.2/58.9 & 19.1/28.7 & 29.6/42.2 & 28.0/38.9 & 30.0/42.2  \\               
    NV-Embed-v2~\cite{lee2025nv} & Tesseract~\cite{smith2007overview} & 36.2/56.0 & 16.9/26.0 & 28.6/44.6 & 30.4/47.8 &  28.0/43.6 \\                
    NV-Embed-v2~\cite{lee2025nv} & GOT-OCR2.0~\cite{wei2024general} & 37.4/53.6 & 20.2/31.5 & 28.7/45.1 & 30.4/43.3 & 29.2/43.4 \\       
    \bottomrule
    \end{tabular}
    }
\end{table*}
\subsection{Additional Experimental Results and Analysis}

\medskip\noindent\textbf{Main results using other metrics.} 
Table~\ref{tab:retrieval_recall} presents the contextual retrieval results on \cmdrdatasetname using Recall@5, in addition to the nDCG@5 results reported in Table~\ref{tab:retrieval_results}. We observe that the Recall@5 results exhibit the same trend as nDCG@5, with our contextual models consistently outperforming the non-contextual baselines.

\medskip\noindent\textbf{Does our model outperform non-contextual models with a reranker?}
For a strong baseline, we apply a state-of-the-art multimodal reranker, Qwen3-VL Reranker~\cite{li2026qwen3}, to the top-10 candidates initially retrieved by the non-contextual model.
As shown in Table~\ref{tab:rerank}, our contextual model still outperforms the non-contextual model even when enhanced with reranking.
While reranking improves the retrieval accuracy of non-contextual models, it introduces additional computational overhead due to the two-stage pipeline.
In contrast, our model incorporates contextual information directly during retrieval, eliminating the need for a separate reranking stage.
Moreover, applying the reranker to our contextual model yields no further improvements. We hypothesize that this is because conventional rerankers are not explicitly trained to capture inter-page dependencies among candidate pages.

\medskip\noindent\textbf{How does OCR quality influence text retriever's performance?} 
We analyze the impact of OCR quality on text-based retrievers by comparing outputs from Tesseract~\cite{smith2007overview} and the state-of-the-art model on Fox~\cite{liu2024focus} benchmark, GOT-OCR2.0~\cite{wei2024general}. 
Table~\ref{tab:ocr} shows that OCR quality has only a limited effect on the final retrieval performance on \cmdrdatasetname. This suggests that retrieval errors are not primarily due to imperfect text recognition, but rather to the inability of text-only retrievers to model cross-page contextual dependencies.

\begin{figure*}
    \centering
\includegraphics[width=1.\textwidth]{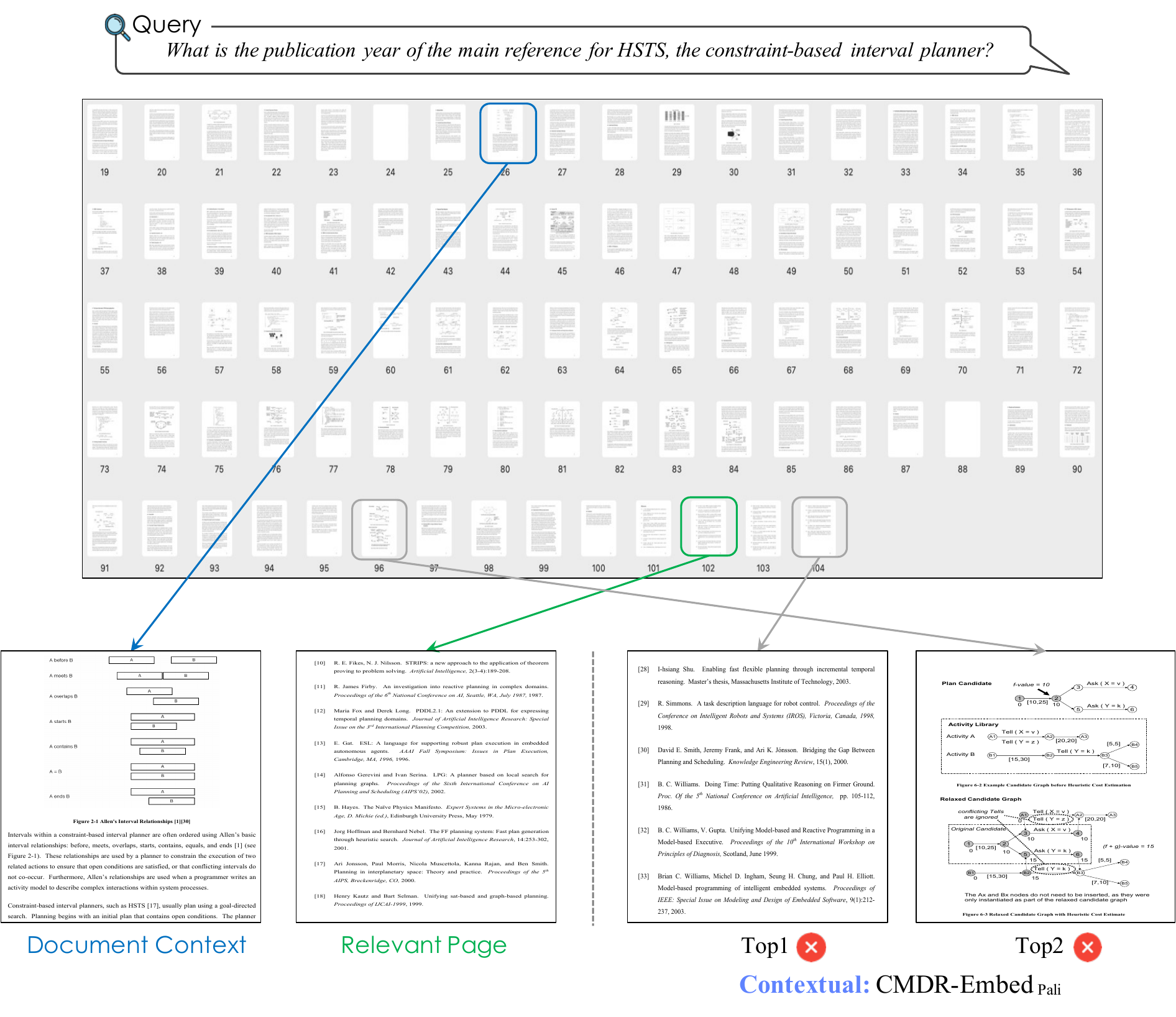}
    \caption{Error example of \textbf
    {Long-context understanding}.}
    \label{fig:error_longcontext}
\end{figure*}

\begin{figure*}
    \centering
\includegraphics[width=1.\textwidth]{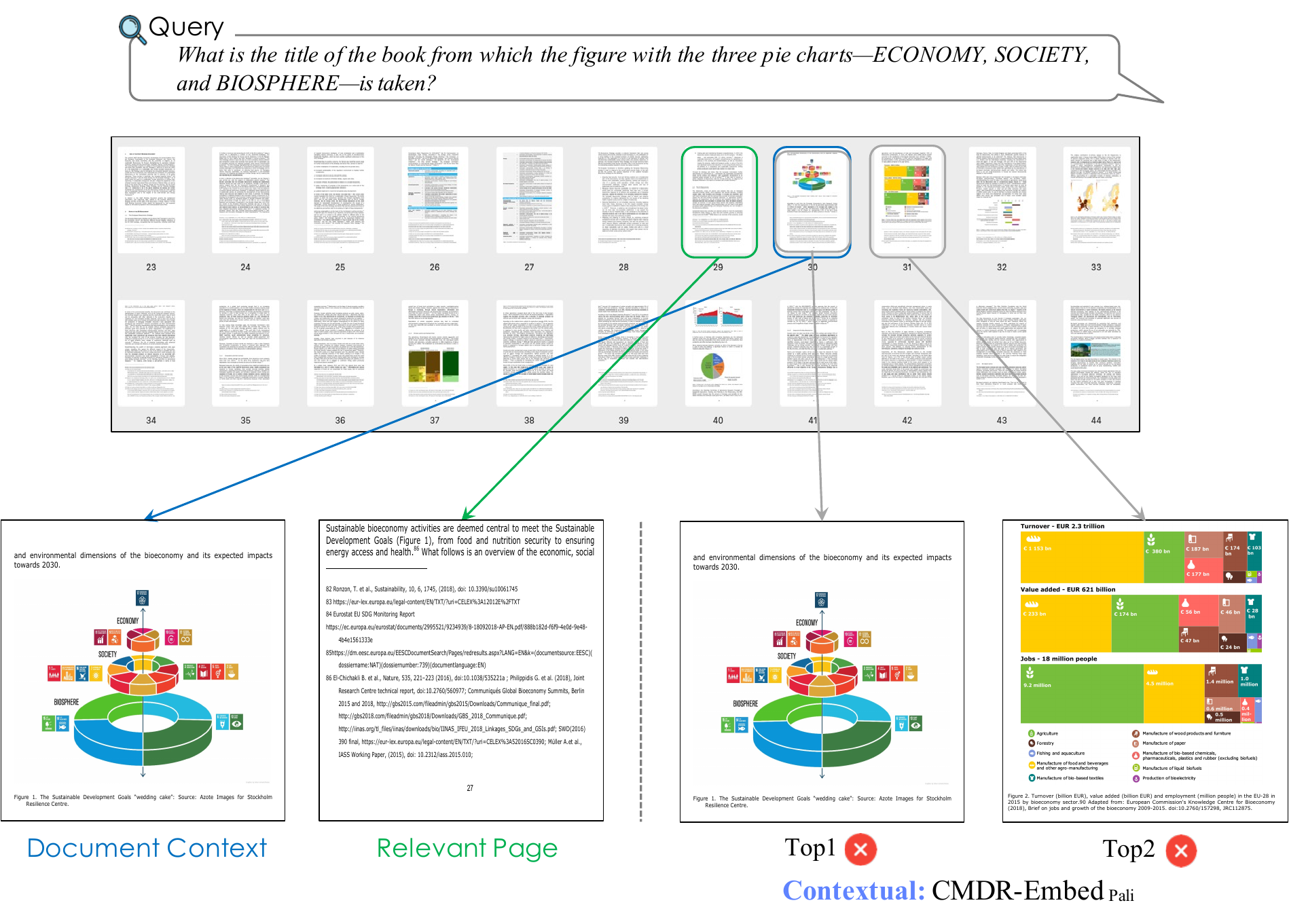}
    \caption{Error example of \textbf
    {Subsequent-context understanding}.}
    \label{fig:error_backwardcontext}
\end{figure*}

\begin{figure*}
    \centering
\includegraphics[width=1.\textwidth]{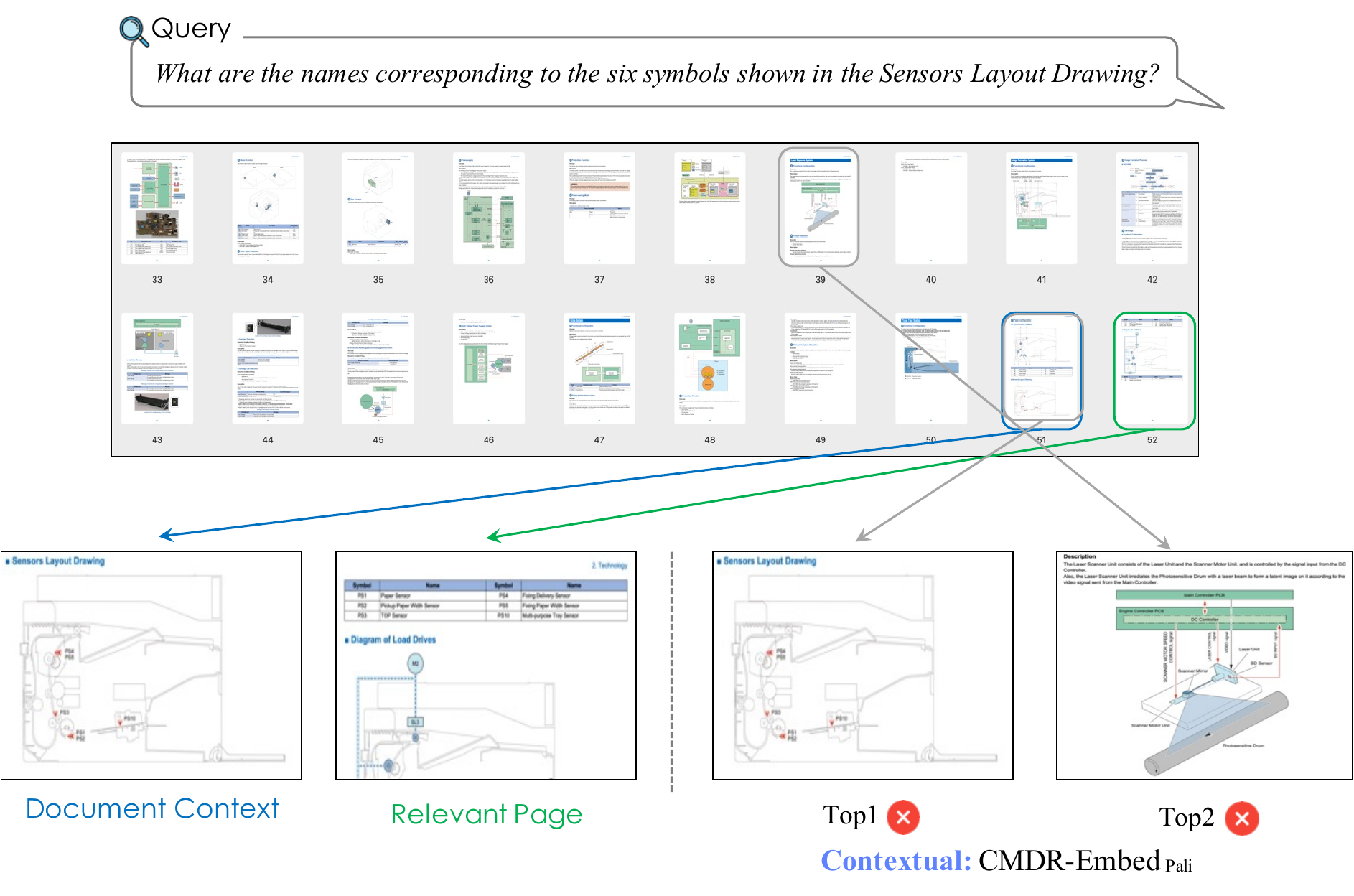}
    \caption{Error example of \textbf
    {Fine-grained visual understanding}.}
    \label{fig:error_visualunderstanding}
\end{figure*}

\medskip\noindent\textbf{Error analysis.}
Figure~\ref{fig:error_longcontext}, Figure~\ref{fig:error_backwardcontext}, and Figure~\ref{fig:error_visualunderstanding} show the error examples of (1) long-context understanding, (2) subsequent-context understanding, and (3) fine-grained visual understanding, respectively.

\end{document}